\documentclass[pra,preprint]{revtex4-1}
\usepackage{hyperref} 

\usepackage{graphicx} 
\usepackage{subfig}
\usepackage{color}
\usepackage{filecontents}
\usepackage{soul}

\newcommand{\ket}[1]{\left| #1 \right\rangle}

\newcommand{\be}{\begin{equation}}
\newcommand{\ee}{\end{equation}}
\newcommand{\bea}{\begin{eqnarray}}
\newcommand{\eea}{\end{eqnarray}}
\newcommand*{\myeqref}[2][Eq.~]{%
\hyperref[{#2}]{#1(\ref*{#2})}%
}
\def\equationautorefname#1#2\null{%
Eq.#1(#2\null)%
}

\usepackage{dcolumn}
\usepackage{bm}
\usepackage{amssymb,amsmath}
\usepackage{color}
\usepackage{float}
\usepackage{tikz}
\usetikzlibrary{arrows}

\definecolor{DarkGreen}{rgb}{0,0.6,0.2}

\begin{document}
\title{Chirality, Band Structure and Localization in Waveguide Quantum Electrodynamics}

\author{Imran M. Mirza}
\affiliation{Department of Physics, University of Michigan, Ann Arbor, Michigan 48109, USA}

\author{Jeremy G. Hoskins}
\affiliation{Department of Mathematics, University of Michigan, Ann Arbor, Michigan 48109, USA}

\author{John C. Schotland}
\affiliation{Department of Mathematics and Department of Physics, University of Michigan, Ann Arbor, Michigan 48109, USA} 

\begin{abstract}
Architectures based on waveguide quantum electrodynamics have emerged as promising candidates for quantum networks. In this paper, we analyze the propagation of single-photons in disordered many-atom waveguides. We pay special attention to the influence of chirality (directionality of photon transport) on the formation of localized photonic states, considering separately the cases of disorder in the atomic positions and in the atomic transition frequencies.
\end{abstract}

\maketitle

\section{Introduction}
The investigation of light-matter interactions in quantum optics is largely concerned with the study of systems consisting of a small number of atoms \cite{gardiner2015quantum}. However, experiments with cold atom systems~\cite{ourjoumtsev2011observation,murmann2015antiferromagnetic}  have led to the study of light propagation in media consisting of a large number of densely-packed scatterers~\cite{javanainen1999one,jennewein2016coherent,zhu2016light}. Moroever,  given the remarkable progress on the scalability of nanophotonic systems in cavity quantum electrodynamics (QED)~\cite{koenderink2015nanophotonics} and ion-trapping techniques~\cite{blatt2012quantum}, the control of quantum states of light coupled to complex atomic media with tunable properties seems to be not far away.

Multi-atom waveguide QED provides a convenient platform to investigate light propagation in complex atomic media. In addition, enhancement of spin-orbit coupling of light in nanoscale waveguides leads to the remarkable ability to control the direction of light propagation~\cite{lodahl2017chiral}. In so-called chiral waveguides, light can propagate preferentially in one direction. Due to this feature, entanglement generation and control \cite{mirza2016multi, mirza2016two}, photon-photon correlations \cite{fang2015waveguide}, superradiance/subradiance \cite{goban2015superradiance}, and selective radiance \cite{asenjo2017exponential} have been extensively investigated. 

Relatively little attention has been paid to the topic of single-photon transport in manybody waveguide QED systems. Shen et al. developed a transfer matrix approach for periodic systems of two-level atoms~\cite{shen2005coherent1}. Witthaut et al. extended this work to the case of three-level atoms and considered the effects of position disorder on single photon transport~\cite{witthaut2010photon}. More recently, Marcuzzi et al. \cite{marcuzzi2017facilitation} investigated position-disordered Rydberg atom systems in tight optical traps. Their experiments were performed for a linear array of up to eight optical tweezers, each containing a single atom, and provided evidence for disorder-induced suppression of excitation transfer. 

In the setting of periodic multi-atom waveguide QED, an important question is to characterize the formation of allowed and forbidden bands for single photon transport. This topic has been addressed for symmetric waveguides~\cite{yanik2004stopping, shen2007stopping, zueco2012microwave}. However, the extent to which chirality can influence band structure and dispersion has not been addressed. In this work, we show that a small chiral imbalance in group velocities can change the location and width of bands compared to symmetric waveguide systems.

There is considerable interest in the study of Anderson localization in photonic systems~\cite{anderson1958absence}. In the setting of waveguide QED, there is a competition between long-ranged waveguide mediated atomic interactions and atomic disorder with short-ranged correlations. In this paper, we consider 
chiral and bidirectional waveguides containing 10--$10^3$ two-level atoms. The effects of two types of disorder are examined: randomness in atomic positions and in atomic transition frequencies. In both cases, we study the single-photon transmission coefficient and localization length as a function of the atom-field detuning and the strength of the disorder. For  chiral waveguides, we find that photon transport is immune to position disorder. However, for frequency disorder localization does occur. For bidirectional waveguides, both types of disorder lead to localization.

The paper is organized as follows. In section II we consider the theory of chiral waveguides and discuss photon transport in both periodic and disordered settings. In Section III, we focus on the non-chiral situation. In Section IV and V, we discuss the band structure and disorder respectively, for both small and symmetric waveguide problems. Finally, in Section VI, we close with a discussion of our results.

\begin{figure}[t]
\includegraphics[width=6.5in,height=1.32in]{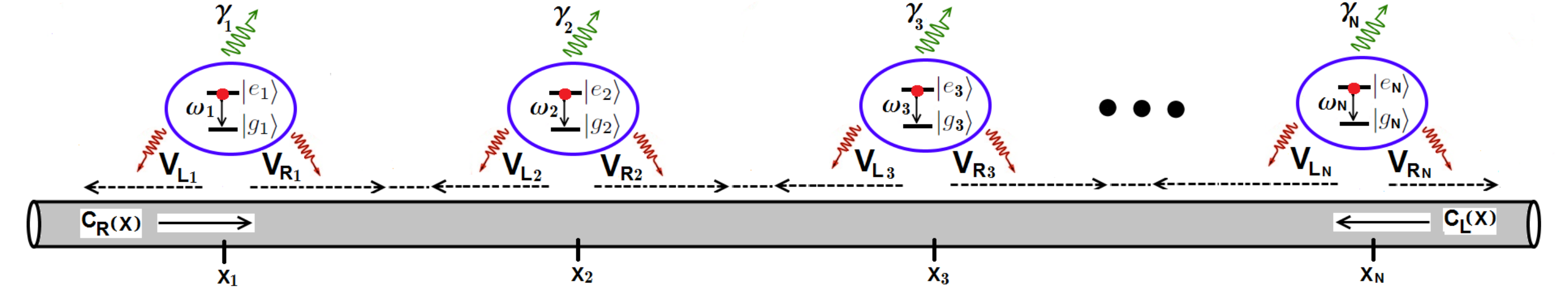}
\captionsetup{
 format=plain,
 margin=1em,
 justification=raggedright,
 singlelinecheck=false
}
\caption{(Color online) Illustrating the waveguide QED system that is considered in this paper.}
\label{Fig1}
\end{figure}

\section{Chiral waveguides}
When confined to subwavelength scales, light shows the remarkable feature of enhanced spin-orbit coupling, enabling the creation of chiral waveguides. Here, chirality is defined as an imbalance in the left and right waveguide emission directions or atom-field coupling strength~\cite{lodahl2017chiral,coles2016chirality,petersen2014chiral,sollner2015deterministic}. In recent years, chiral waveguide QED has undergone tremendous development~\cite{mitsch2014quantum, coles2016chirality} in which up to 90\% directionality has been reported. We note that chiral waveguides are similar to a waveguide-coupled ring resonators, in which light propagation is also unidirectional~\cite{hach2010fully}.

In this paper we consider the following scenario for both chiral and bidirectional waveguides. A collection of $N$ two-level atoms (also referred to as qubits or emitters) is side-coupled to a one-dimensional lossless and dispersionless waveguide, as shown in Fig.\ref{Fig1}. This model can be experimentally realized in a number of different physical settings including cadmium selenide quantum dots interacting with silver nanowires~\cite{akimov2007generation}, Josephson junctions in microwave transmission lines~\cite{astafiev2010resonance}, Cesium atoms coupled to photonic crystal waveguide~\cite{goban2015superradiance}, and silicon-vacancy color centers coupled to diamond nanowaveguides~\cite{sipahigil2016integrated}.

We consider the following Hamiltonian for a multiatom chiral waveguide system:
\begin{equation}
\label{eq:HsysC}
\begin{split}
\hat{H}=&\sum_{j}(\omega_{{j}}-i\gamma_{j})\hat{\sigma}^{\dagger}_{j}\hat{\sigma}_{j}+\int dx\hat{c}^{\dagger}(x)\left(\omega_{0}-iv_{g}\frac{\partial}{\partial x}\right)\hat{c}(x)\\
&+ \sum_{j}\int dx\delta(x-x_{j})\left[V_{j}\hat{c}^{\dagger}(x)\hat{\sigma}_{j}+{h.c.}\right] \ .
\end{split}
\end{equation}
The first term in (\ref{eq:HsysC}) corresponds to the Hamiltonian of the atoms, the second term to the Hamiltonian of the quantized field, and the third term to the interaction between the atoms and the field. Here we work in units where $\hbar=1$, have employed the method of real-space quantization~\cite{shen2005coherent1,shen2005coherent} and have made the rotating wave approximation. The position of the $j$th atom is denoted $x_j$ and its transition frequency is $\omega_j$ with $j=1,\ldots , N$. In addition, $\omega_{0}$ is the frequency around which waveguide dispersion relation has been linearized, $v_{g}$ is the group velocity of the photon in the waveguide and $\gamma_{j}$ is the rate of spontaneous emission of the $j$th atom.  The atomic lowering operator is denoted $\hat{\sigma}_{j}$ and the field operator $\hat{c}(x)$ annihilates a photon at the position $x$. The nonvanishing commutation relations are given by
\begin{eqnarray}
\left[\hat{c}(x),\hat{c}^\dag(x')\right] = \delta(x-x') \ , \quad \{\hat{\sigma}_{i}, \hat{\sigma}_{j}^\dag \} = \delta_{ij} \ .
\end{eqnarray}
Finally,  $V_j$ is the evanescent coupling of the atom to the waveguide continuum. 

The quantum state of the system in the subspace of zero and one excitations is of the form
\begin{equation}
\label{eq:Psi(t)}
\begin{split}
&\ket{\Psi}=\int dx \varphi(x)\hat{c}^{\dagger}(x)\ket{\varnothing}+\sum_{j}a_{j}\hat{\sigma}^{\dagger}_{j}\ket{\varnothing} \ ,
\end{split}
\end{equation} 
where $a_{j}$ is the probability amplitude for the $j$th atom, $\varphi(x)$ is the one-photon amplitude and $\ket{\varnothing}$ is the ground state of the atom-waveguide system. 
The equations obeyed by  $a$ and $\varphi$ can be obtained by substituting the above expression for $\ket{\Psi}$ into the time-independent Schr\"odinger equation $\hat{H}\ket{\Psi}=\hbar\omega\ket{\Psi}$, where $\omega$ is the frequency of the photon. We thus obtain
\begin{subequations}
\begin{eqnarray}
\label{eq:TIAE11}
-iv_{g}\frac{\partial \varphi(x)}{\partial x}+\sum^{N}_{j=1}V_{j}a_{j}\delta(x-x_{j})&=&\left(\omega-\omega_0\right)\varphi(x) \ , \\
V^{\ast}_{j}\varphi(x_{j})&=&(\omega-\omega_{{j}}+i\gamma_{j})a_{j} \ .
\end{eqnarray}
\end{subequations}
{Eliminating $a_j$ from  (\ref{eq:TIAE11}) yields the following equation for $\varphi$:}
{\begin{equation}
\label{eq:phiEq}
-iv_{g}\frac{\partial \varphi(x)}{\partial x}+\sum^{N}_{j=1}v_j\delta(x-x_{j}) \varphi(x)
=(\omega-\omega_{0})\varphi(x) ,
\end{equation}
where $v_j =|V_{j}|^{2}/ \left(\omega-\omega_{{j}}-i\gamma_{j}\right)$.
The solution to (\ref{eq:phiEq}) can be obtained by observing that in between the atoms, when $x\neq x_j$,   $\varphi(x)=Ae^{iqx}$, where the wavenumber $q=(\omega-\omega_{0})/v_g$ and $A$ is constant. Thus $\varphi$ is of the form
\begin{equation}
\label{eq:phi}
\varphi(x)=
\begin{cases}
  e^{iqx}, \hspace{5mm}x<x_1,\\      
  t_{1}e^{iqx}, \hspace{2mm} x_{1} \le x \le x_2, \\
  \vdots \\
  t_{N}e^{iqx}, \hspace{2mm}x>x_N  .
\end{cases}
\end{equation}
To obtain the coefficients $t_j$, we integrate (\ref{eq:phiEq}) over the interval $[x_j-\epsilon,x_j+\epsilon]$, where $\epsilon$ is a small positive number. This yields the jump condition
\begin{equation}
\label{eq:boundaryC}
iv_{g}[\varphi(x_{j}+\epsilon)-\varphi(x_{j}-\epsilon)]=v_j\varphi(x_{j}) .
\end{equation}
Next, we regularize  the discontinuity in $\varphi$ according to
\begin{equation}
\varphi(x_{j})=\lim_{\epsilon\longrightarrow 0}\left[\varphi(x_{j}+\epsilon)+\varphi(x_{j}-\epsilon)\right]/2
\end{equation}
and introduce the quantity $\Gamma_{j}={|V_{j}|^{2}}/{2v_{g}}$. Eq.~(\ref{eq:boundaryC}) thus becomes
{\begin{equation}
\varphi(x_{j}+\epsilon)=T_j\varphi(x_{j}-\epsilon) ,
\end{equation}}
where 
\begin{equation}
\label{def_Tj}
T_j = \frac{\omega-\omega_{{j}}+i(\gamma_{j}-\Gamma_{j})}{\omega-\omega_{{j}}+i(\gamma_{j}+\Gamma_{j})} .
\end{equation} 
Finally, {by using (\ref{eq:phi}) we arrive at the recursion relation
\begin{equation}
\label{eq:Ct&t}
t_{j}= T_j t_{j-1} ,
\end{equation}
which allows us to determine the amplitude $\varphi$.

To study the transport of single photons, we define the transmission coefficient $T=|\varphi(x_N)/\varphi(x_1)|^2$, which upon making use of (\ref{eq:phi}) and (\ref{eq:Ct&t}) becomes 
\begin{equation}
\label{eq:netTC}
T = \prod_{j=1}^N |T_j|^2 \ .
\end{equation}
As expected, if $\gamma_{j}=0$ (no losses), then $T=1$ and the system behaves as an all-pass filter. 

\subsection{Periodic arrangement}
Eq.~(\ref{eq:Ct&t}) is applicable to both periodic and disordered arrangements of atoms. In the periodic case, unlike the bidirectional setting discussed in section~III, there is no band structure and the transmission is independent of the period. On the other hand, it is convenient to distinguish three regimes when 
$\gamma_{j}\neq 0$: undercoupled ($\gamma_{j}>\Gamma_{j}$), overcoupled ($\gamma_{j}<\Gamma_{j}$), and critically coupled ($\gamma_{j}=\Gamma_{j}$).
\begin{figure}
\centering
   \begin{tabular}{@{}cccc@{}}
   \includegraphics[width=2.9in, height=2in]{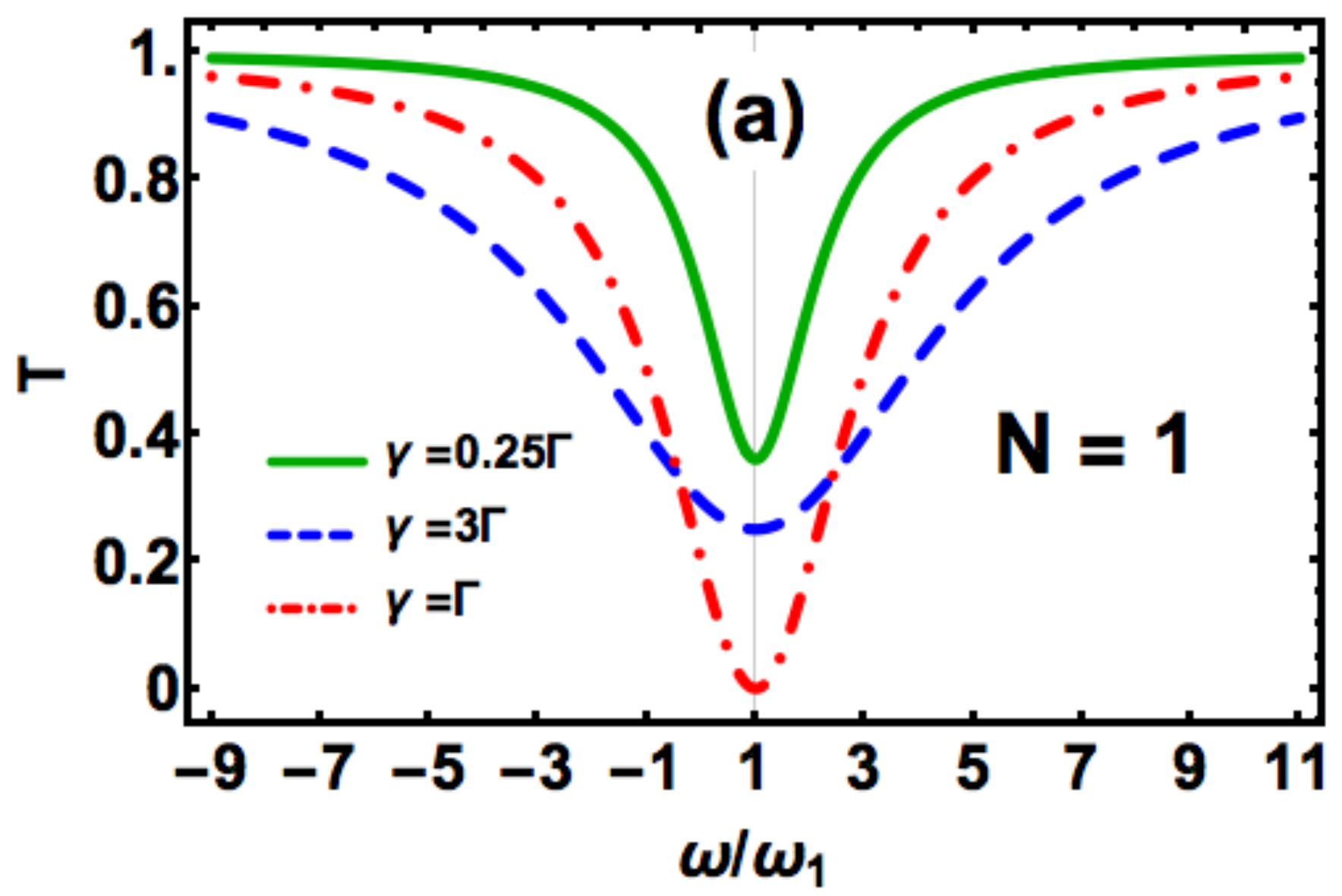} 
  \hspace{4mm}\includegraphics[width=2.9in, height=2in]{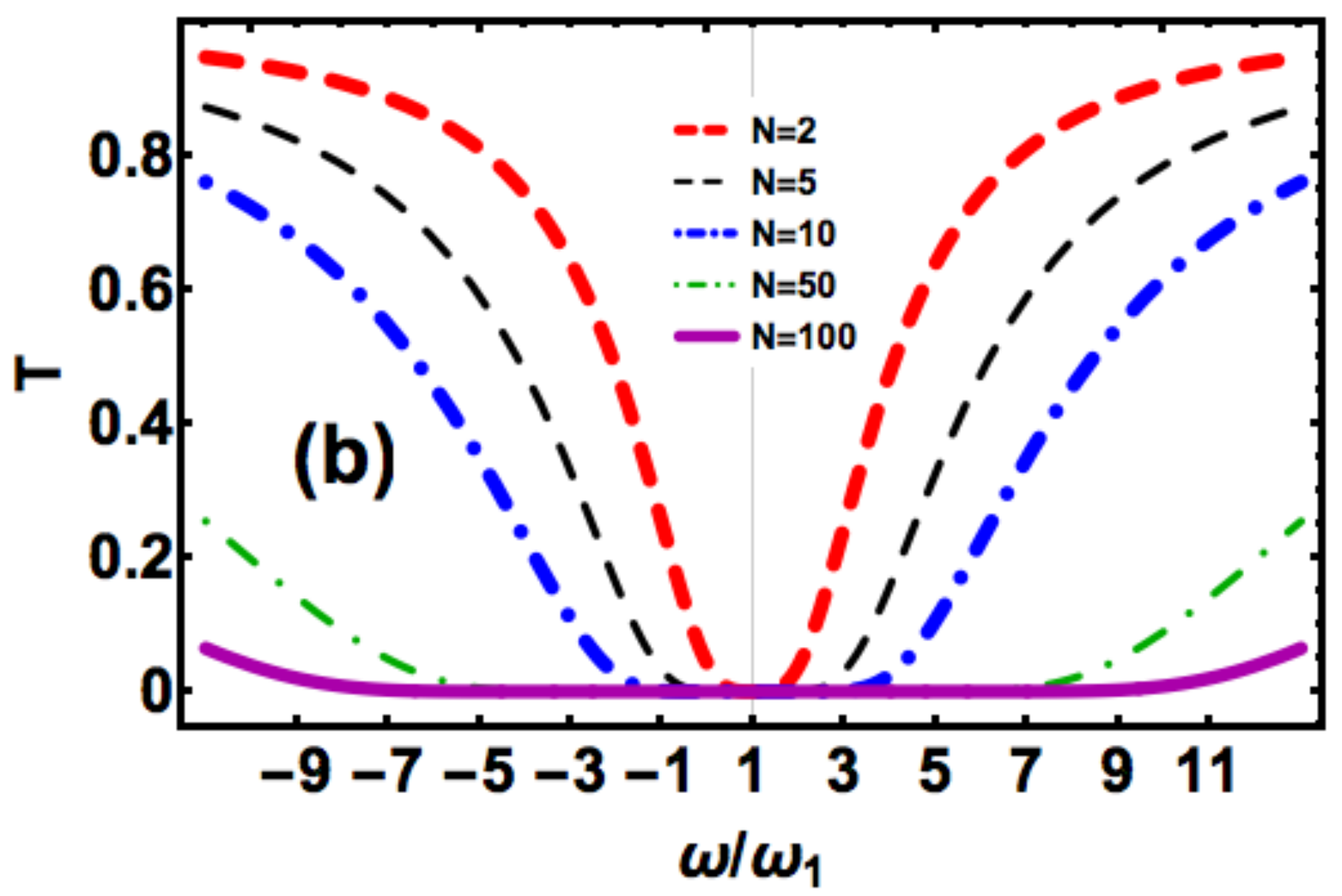}
  \end{tabular} 
\captionsetup{
  format=plain,
  margin=1em,
  justification=raggedright,
  singlelinecheck=false}
\caption{(Color online) Transmission of a single photon in a chiral system consisting of {(a)} 1 and {(b)} 2, 5, 10, 50 and 100 periodically arranged identical atoms ($\omega_{j}=\omega_{1}$ and $\gamma_{j}=\gamma$ for all $j$). In (a) the solid green, red dotted-dashed and blue dashed lines represent the over-, critical- and under-coupled regimes, respectively. In (b) we have chosen the critical coupling case with $\gamma=\Gamma$.}
\label{Fig2}
\end{figure}
As is evident from the single atom case, which we show in Fig.~\ref{Fig2}(a), in the critical coupling regime the transmission reaches its minimum value. In Fig.~\ref{Fig2}(b), we plot the transmission for different numbers of identical atoms in the critical coupling regime. As we increase the number of atoms, we notice the width of the region of low transmission grows and for a 100-atom chain, transmission is suppressed for a wide range of frequencies. 

\subsection{Disordered arrangement}
We now introduce disorder in the multi-atom chain and investigate the occurence of single photon localization. For recent studies on localization in photonic systems, see for instance \cite{segev2013anderson,wiersma2013disordered,javadi2014statistical,joannopoulos2011photonic,crespi2013anderson}. In what follows and for the rest of the paper, all random variables are generated from a Gaussian probability density of the form
\begin{equation}
\label{gaussian}
P(x)=\frac{1}{\sqrt{2\pi\sigma^2}}e^{-(x-\overline{x})^{2}/2\sigma^{2}},
\end{equation}
where $\overline{x}$ is the mean and $\sigma$ being the standard deviation is a measure of the strength of the disorder.

\subsubsection{Frequency disorder}
Here we consider the case of frequency disorder, in which we assume that the atomic transition frequencies are random. This type of disorder can be present in optically trapped Rydberg atoms, either due to non-uniformity of the applied potential or when beam focusing is inhomogeneous~\cite{zhang2011magic,maller2015rydberg}. We begin by calculating the average transmission and then compute the localization length.

Suppose that the detunings $\delta_j = \omega-\omega_j$ are independent and identically distributed Gaussian random variables. Making use of (\ref{def_Tj}) and (\ref{eq:Ct&t}), we find that the average transmission for an $N$-atom chain is given by
\begin{eqnarray}
\langle T \rangle &=& \int \prod_{j=1}^N d\delta_j P(\delta_j) |T_j|^2 \\
\label{TotalTC}&=& \langle |\tau|^2 \rangle^N.
\label{avgT}
\end{eqnarray}
Here
\begin{equation}
\langle |\tau|^2 \rangle = \int d\delta P(\delta) |\tau|^2  ,
\end{equation}
where 
\begin{equation}
\tau = \frac{\delta + i (\gamma-\Gamma)}{\delta + i (\gamma+\Gamma)} .
\end{equation}
It is easily seen that
\begin{equation}
\label{tau_squared}
|\tau|^{2}=1-\left[\left(\gamma+\Gamma\right)^{2}-\left(\gamma-\Gamma\right)^{2}\right]\int^{\infty}_{0}e^{-\lambda\left(\delta^{2}+(\gamma+\Gamma)^{2}\right)}d\lambda .
\end{equation}
Carrying out the indicated average over $\delta$ with $\langle\delta\rangle=\bar\delta$ yields
\begin{equation}
\langle |\tau|^2 \rangle = 1-4\gamma\Gamma\int^{\infty}_{0}\frac{\exp\left[{-\lambda(\gamma+\Gamma)^{2}-\frac{\lambda\overline{\delta}^{2}}{1+2\lambda\sigma^{2}}}\right]}{\sqrt{1+2\lambda\sigma^{2}}}d\lambda ,
\end{equation}
which allows us to calculate the average transmission from (\ref{avgT}) .

By analogy to the theory of disordered electronic systems~\cite{markos2008wave,izrailev1999localization,delande2013many}, we define the localization length $\xi$ by
\begin{equation}
\label{llc}
\xi^{-1} = - \lim_{N\to\infty} \frac{\langle \ln T \rangle}{N} \ ,
\end{equation} 
where the average is over all detunings $\delta_j$. It is easily seen from (\ref{eq:netTC}) and (\ref{tau_squared}) that
\begin{equation}
\langle \ln T \rangle = N \langle \ln |\tau|^2 \rangle
\end{equation}
and thus
\begin{equation}
\xi^{-1} = - \langle \ln |\tau|^2 \rangle .
\end{equation}
In  the critical coupling regime with $\gamma=\Gamma$, we can perform the above average explicitly and thus obtain
\begin{equation}
\begin{split}
&\xi^{-1}=-\frac{2\Gamma}{\sqrt{2\pi\sigma^2}}\int^{\infty}_{-\infty}\ln\left(1-\frac{1}{1+x^{2}}\right)e^{-(2\Gamma x-\overline{\delta})^{2}/2\sigma^{2}}dx.
\end{split}
\end{equation}
\begin{figure}
\centering
   \begin{tabular}{@{}cccc@{}}
   \includegraphics[width=3.1in, height=2in]{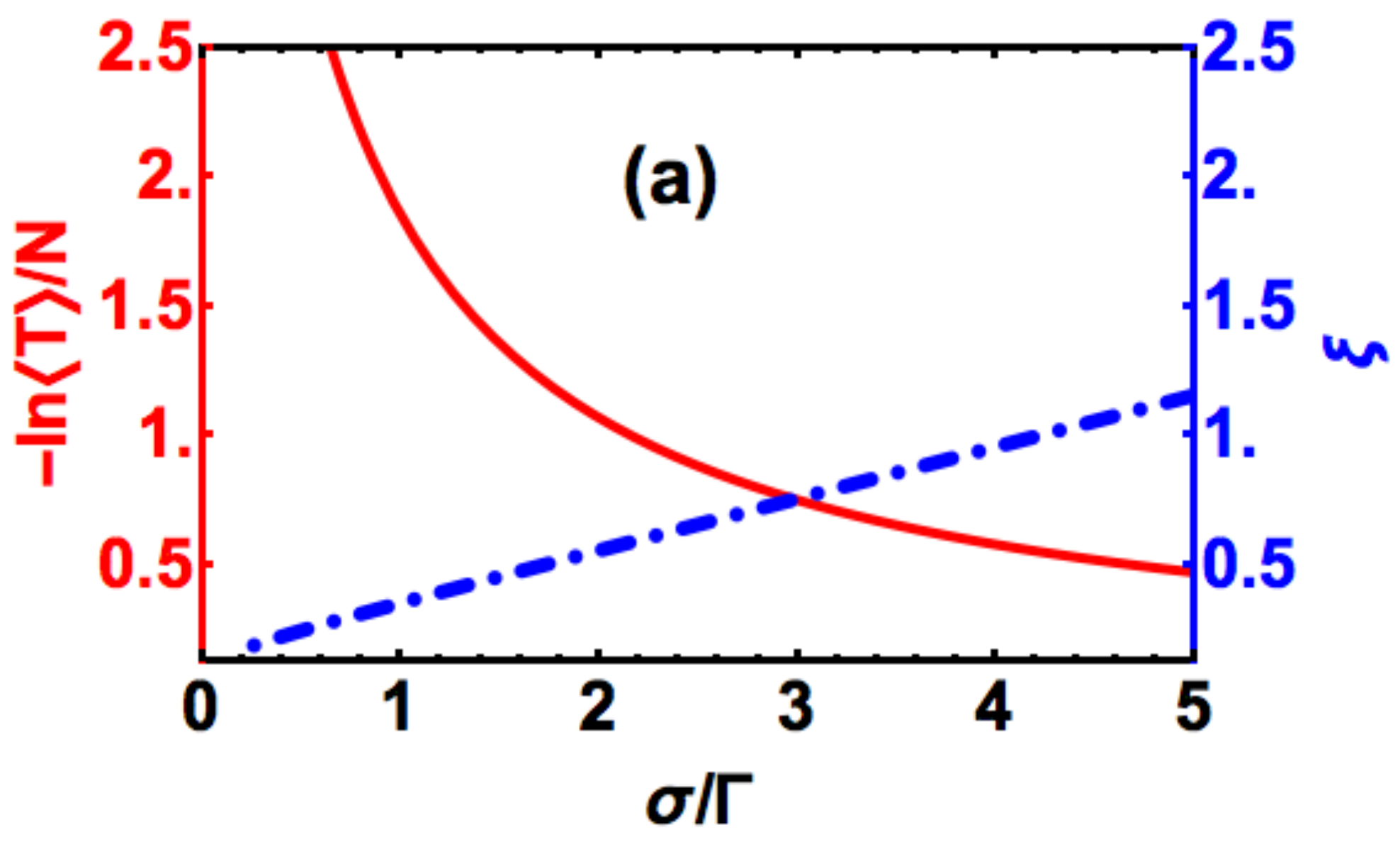} 
  \hspace{4mm}\includegraphics[width=3.1in, height=2in]{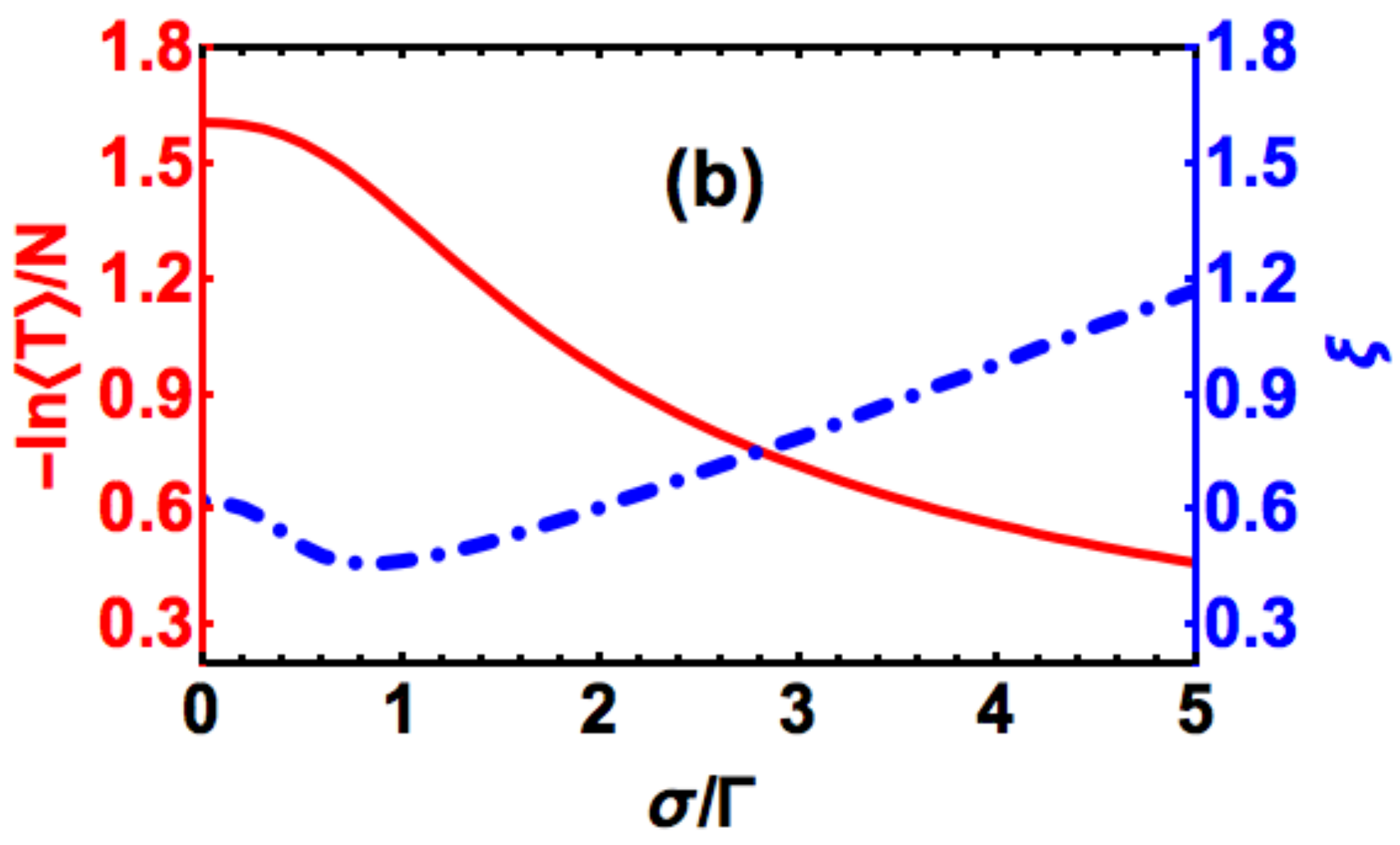}
  \end{tabular} 
\captionsetup{
  format=plain,
  margin=1em,
  justification=raggedright,
  singlelinecheck=false}
\caption{(Color online) Average transmission and localization length for a frequency-disordered chiral waveguide in the critical coupling regime. (a) $\bar\delta = 0$ and (b) $\bar\delta=\Gamma$.}
\label{Fig3}
\end{figure}

In Fig.~\ref{Fig3}, we plot the average transmission and localization length as a function of the strength of the disorder $\sigma$ and the average detuning $\bar\delta$. We first consider the case $\bar\delta=0$. We find that the system is purely reflecting ($\langle T\rangle=0$) 
when $\sigma=0$. This is a consequence of the fact that the system is both on resonance and critically coupled. 
We note that as $\sigma$ is  increased the transmission increases, as does the localization length. Next we consider the case $\bar\delta=\Gamma$. Here we see that even when $\sigma=0$, the system is off resonance and the average transmission is nonvanishing. Note the presence of a minimum in the localization length near $\sigma=\Gamma$.

\subsubsection{Position disorder}

We now consider the effect of position disorder of the atoms in the chain. It follows immediately from (\ref{eq:netTC})
that the transmission $T$ does not depend on the position of the atoms. Thus in chiral waveguides we see that transport is immune to position disorder.

\section{Bi-directional waveguides} 
We consider the following Hamiltonian for a multiatom bidirectional waveguide
\begin{equation}
\label{eq:Hsys}
\begin{split}
\hat{H}=&\sum_{j}(\omega_{j}-i\gamma_{j})\hat{\sigma}^{\dagger}_{j}\hat{\sigma}_{j}+\int dx\hat{c}^{\dagger}_{R}(x)\left(\omega_{0} - iv_{R}\frac{\partial}{\partial x}\right)\hat{c}_{R}(x)  + \int dx\hat{c}^{\dagger}_{L}(x)\left(\omega_{0} + iv_{L}\frac{\partial}{\partial x}\right)\hat{c}_{L}(x)\\
&+\sum_{m,j}\int dx\delta(x-x_{j})\left[V_{mj}\hat{c}^{\dagger}_{m}(x)\hat{\sigma}_{j}+h.c.\right] .
\end{split}
\end{equation}
The first term in (\ref{eq:Hsys}) is the Hamiltonian of the atoms. The second and third terms are the Hamiltonian of the waveguide, which supports left- and right-going modes with group velocities $v_R$ and $v_L$, respectively. Here the the sum is over $m\in\{R,L\}$. The destruction of a single photon in the left (right) waveguide continuum at position $x$ is represented by the field operator $\hat{c}_{L}(x)(\hat{c}_{R}(x))$. The nonvanishing commutation relations for field operators are given by 
\begin{eqnarray}
\left[\hat{c}_m(x),\hat{c}^\dag_{n}(x')\right] = \delta_{mn}\delta(x-x').
\end{eqnarray}
The third term in (\ref{eq:Hsys}) accounts for the interaction between the quantized field and the atoms, with $V_{mj}$ denoting the corresponding coupling, which is chosen to be real-valued. The waveguide described by the Hamiltonian (\ref{eq:Hsys}) is said to be \emph{bidirectional}. If $v_R=v_L$ and $V_{Rj}=V_{Lj}$ the waveguide  is referred to as \emph{symmetric}. Evidently, the extreme case with either $v_R$ or $v_L$ vanishing corresponds to a chiral waveguide. 

We consider a one-photon quantum state of the form
\begin{equation}
\label{eq:Psi(t)}
\begin{split}
&\ket{\Psi}=\sum_{m}\int dx \varphi_{m}(x)\hat{c}^{\dagger}_{m}(x)\ket{\varnothing}+\sum_{j}a_{j}\hat{\sigma}^{\dagger}_{j}\ket{\varnothing}.
\end{split}
\end{equation} 
Here $\varphi_{R}(x),(\varphi_{L}(x))$ is the one-photon amplitude in the right (left) waveguide continuum. As in section II, the equations obeyed by the amplitudes can be obtained from the Schr\"odinger equation. We thus obtain
\begin{subequations}
\label{eq:TIAENC}
\begin{eqnarray}
-iv_{R}\frac{\partial \varphi_{R}(x)}{\partial x}+\sum^{N}_{j=1}V_{R_{j}}a_{j}\delta(x-x_{j})=(\omega-\omega_0)\varphi_{R}(x) , \\
iv_{L}\frac{\partial \varphi_{L}(x)}{\partial x}+\sum^{N}_{j=1}V_{L_{j}}a_{j}\delta(x-x_{j})=(\omega-\omega_0)\varphi_{L}(x) , \\ 
V_{R_{j}}\varphi_{R}(x_{j})+V_{L_{j}}\varphi_{L}(x_{j})=(\omega-\omega_{j}+i\gamma_{j})a_{j}.
\end{eqnarray}
\end{subequations}
Eliminating $a_{j}$ from the above, we find that the following equations are obeyed by $\varphi_{R}$ and $\varphi_{L}$:
\begin{subequations}
\label{eq:TIAENC1}
\begin{eqnarray}
-iv_{R}\frac{\partial \varphi_{R}(x)}{\partial x}+\sum_{j}\frac{V_{R_{j}}\delta(x-x_{j})}{\omega-\omega_{j}+i\gamma_{j}}\Bigg(V_{R_{j}}\varphi_{R}(x)+V_{L_{j}}\varphi_{L}(x)\Bigg)=(\omega-\omega_0)\varphi_{R}(x), \hspace{10mm}\\
iv_{L}\frac{\partial \varphi_{L}(x)}{\partial x}+\sum_{j}\frac{V_{L_{j}}\delta(x-x_{j})}{\omega-\omega_{j}+i\gamma_{j}}
\Bigg(V_{R_{j}}\varphi_{R}(x)+V_{L_{j}}\varphi_{L}(x)\Bigg)
=(\omega-\omega_0)\varphi_{L}(x), \hspace{10mm}
\end{eqnarray}
\end{subequations}

\begin{figure}[t]
\includegraphics[width=6.6in,height=1.05in]{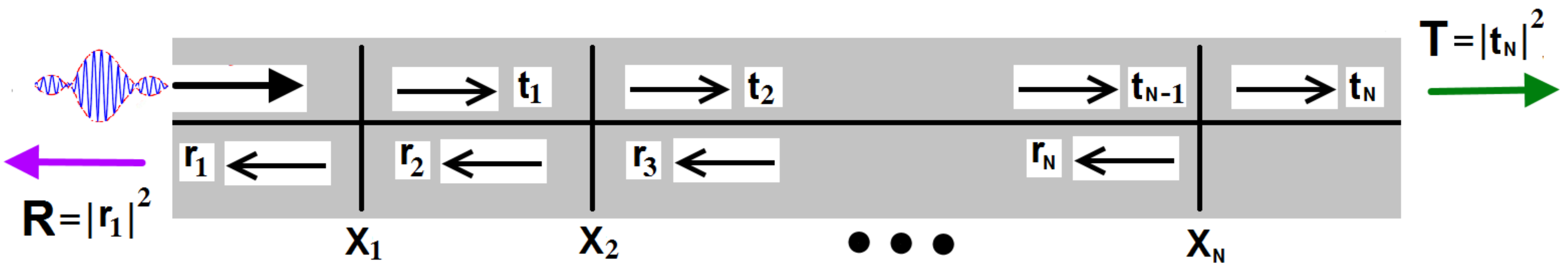}
\captionsetup{
 format=plain,
 margin=1em,
 justification=raggedright,
 singlelinecheck=false
}
\caption{(Color online) Illustrating the transmission and reflection amplitudes at the location of each atom.}
\label{Fig4}
\end{figure}

\noindent
When $x\neq x_{j}$, the amplitudes $\varphi_{R}(x)$ and $\varphi_{L}(x)$ are given by
$\varphi_{R}(x)=A_Re^{iq_{R}x}$ and $\varphi_{L}(x)=A_Le^{-iq_{L}x}$. Here  $q_{R}=(\omega-\omega_{0})/v_{R}$, $q_{L}=(\omega-\omega_{0})/v_{L}$ are the wavenumbers associated with the right and left field amplitudes, respectively, and $A_R$ and $A_L$ are constant. Thus, we obtain\begin{equation}
\label{eq:phiR}
\varphi_{R}(x)=
\begin{cases}
  e^{iq_{R}x}, \hspace{5mm}x<x_1,\\      
  t_{1}e^{iq_{R}x}, \hspace{2mm} x_{1} \le x \le x_2, \\
  \vdots \\
  t_{N}e^{iq_{R}x}, \hspace{2mm}x>x_N  .
\end{cases}
\end{equation}
and
\begin{equation}
\label{eq:phiL}
\varphi_{L}(x)=
\begin{cases}
  r_{1}e^{-iq_{L}x}, \hspace{5mm}x<x_1,\\      
  r_{2}e^{-iq_{L}x}, \hspace{2mm} x_{1} \le x \le x_2, \\
  \vdots \\
   r_{N}e^{-iq_{L}x}, \hspace{2mm} x_{N-1} \le x \le x_N, \\
  0, \hspace{2mm}x>x_N  .
\end{cases}
\end{equation}
where $r_{N+1}=0$ and $t_{0}=1$. See Fig.~\ref{Fig4}. In order to obtain the coefficients $t_{j}$ and $r_{j}$ we integrate (\ref{eq:TIAENC1}) over the interval $[x_{j}-\epsilon,x_{j}+\epsilon]$, which yields the jump conditions
\begin{subequations}
\begin{eqnarray}
-iv_{R}\Bigg[\varphi_{R}(x_{j}+\epsilon)-\varphi_{R}(x_{j}-\epsilon)\Bigg]+\frac{V_{R_{j}}}{\omega-\omega_{j}+i\gamma_{j}}\Bigg(V_{R_{j}}\varphi_{R}(x_{j})+V_{L_{j}}\varphi_{L}(x_{j}) \Bigg)=0,\hspace{5mm}\\
iv_{L}\Bigg[\varphi_{L}(x_{j}+\epsilon)-\varphi_{L}(x_{j}-\epsilon)\Bigg]+\frac{V_{L_{j}}}{\omega-\omega_{j}+i\gamma_{j}}\Bigg(V_{L_{j}}\varphi_{L}(x_{j})+V_{R_{j}}\varphi_{R}(x_{j}) \Bigg)=0.\hspace{5mm}
\end{eqnarray}
\end{subequations}
Regularizing $\varphi_{m}$ by
\begin{equation}
\varphi_{m}(x)=\lim_{\epsilon\longrightarrow 0}\left[\varphi_{m}(x_{j}+\epsilon)+\varphi_{m}(x_{j}-\epsilon)\right]/2, 
\end{equation}
and introducing the quantities $\Gamma_{R_{j}}=V_{R_{j}}^{2}/2v_{R}$ and $\Gamma_{L_{j}}={V_{L_{j}}^{2}}/{2v_{L}}$, we obtain 
\begin{subequations}
\begin{eqnarray}
\varphi_{R}(x_{j}+\epsilon)=\Bigg(\frac{{\Delta}_{j}-i\Gamma_{R_{j}}}{{\Delta}_{j}+i\Gamma_{R_{j}}} \Bigg)\varphi_{R}(x_{j}-\epsilon)-i\sqrt{\frac{v_{L}}{v_{R}}}\frac{\sqrt{\Gamma_{R_{j}}\Gamma_{L_{j}}}}{{\Delta}_{j}+i\Gamma_{R_{j}}}\Bigg(\varphi_{L}(x_{j}+\epsilon)+\varphi_{L}(x_{j}-\epsilon)\Bigg),\hspace{8mm}\\
\varphi_{L}(x_{j}+\epsilon)=\Bigg(\frac{{\Delta}_{j}+i\Gamma_{L_{j}}}{{\Delta}_{j}-i\Gamma_{L_{j}}} \Bigg)\varphi_{L}(x_{j}-\epsilon)+i\sqrt{\frac{v_{R}}{v_{L}}}\frac{\sqrt{\Gamma_{L_{j}}\Gamma_{R_{j}}}}{{\Delta}_{j}-i\Gamma_{L_{j}}}\Bigg(\varphi_{R}(x_{j}+\epsilon)+\varphi_{R}(x_{j}-\epsilon)\Bigg) , \hspace{10mm}
\end{eqnarray}
\end{subequations}
where ${\Delta}_{j}=\omega-\omega_{j}-i\gamma_{j}$. Using Eq.~(\ref{eq:phiR}) and (\ref{eq:phiL}) we obtain the recursion relations
\begin{subequations}
\label{eq:TIAENC2}
\begin{eqnarray}
t_{j}=\Bigg(\frac{{\Delta}_{j}-i\Gamma_{R_{j}}}{{\Delta}_{j}+i\Gamma_{R_{j}}} \Bigg)t_{j-1}-i\sqrt{\frac{v_{L}}{v_{R}}}\frac{\sqrt{\Gamma_{R_{j}}\Gamma_{L_{j}}}}{{\Delta}_{j}+i\Gamma_{R_{j}}}\Bigg(r_{j}e^{-i(q_{R}+q_{L})x_{j}}+r_{j+1}e^{-i(q_{R}+q_{L})x_{j}}\Bigg),\\
r_{j+1}=\Bigg(\frac{{\Delta}_{j}+i\Gamma_{L_{j}}}{{\Delta}_{j}-i\Gamma_{L_{j}}} \Bigg)r_{j}+i\sqrt{\frac{v_{R}}{v_{L}}}\frac{\sqrt{\Gamma_{L_{j}}\Gamma_{R_{j}}}}{{\Delta}_{j}-i\Gamma_{L_{j}}}\Bigg(t_{j}e^{i(q_{R}+q_{L})x_{j}}+t_{j-1}e^{i(q_{R}+q_{L})x_{j}}\Bigg).
\end{eqnarray}
\end{subequations}
Next, we write transmission and reflection coefficients in terms of the free propagation phase accumulated by the photon as it propagates through the waveguide between two consecutive emitters:
\begin{equation}
t_{j}=\widetilde{t}_{j}e^{-i(q_{R}+q_{L})x_{j}/2},\hspace{5mm} r_{j}=\widetilde{r}_{j}e^{i(q_{R}+q_{L})x_{j-1}/2},
\end{equation}
which defines $\widetilde{r}_{j}$ and $\widetilde{t}_{j}$.
After some rearrangement, (\ref{eq:TIAENC2}) can be expressed in the form of the matrix recursion relation
\begin{equation}
\label{eq:TransEq}
\begin{pmatrix}
    \widetilde{t}_{j}     \\
    \widetilde{r}_{j+1}  \\
\end{pmatrix}=\mathcal{T}_{j}\begin{pmatrix}
\widetilde{t}_{j-1}\\
\widetilde{r}_{j}\\
\end{pmatrix}.
\end{equation} 
Here the transfer matrix $\mathcal{T}_{j}$ is given by
\begin{equation}
\label{eq:T1B}
\mathcal{T}_{j}=
\begin{pmatrix}
    e^{i\phi_{j}}/s^{\ast}_{j}       & -p^{\ast}_{j}/s^{\ast}_{j} e^{-i\phi_{j}} \\
    -p_{j}e^{i\phi_{j}}/s_{j}      & e^{-i\phi_{j}}/s_{j}  \\
\end{pmatrix},
\end{equation} 
where
\begin{equation}
s_{j}=\frac{{\Delta}_{j}-i(\Gamma_{R_{j}}-\Gamma_{L_{j}})}{{\Delta}_{j}+i(\Gamma_{R_{j}}+\Gamma_{L_{j}})},\\
\hspace{3mm}p_{j}=\frac{-2i\sqrt{\Gamma_{R_{j}}\Gamma_{L_{j}}}}{{\Delta}_{j}+i(\Gamma_{R_{j}}+\Gamma_{L_{j}})},\hspace{3mm}\\
\hspace{3mm}\phi_{j}=(q_{R}+q_{L})(x_{j}-x_{j-1})/2 .
\end{equation}
Note that using the above transfer matrix formalism, we recover the results of Shen and Fan \cite{fan2002sharp, shen2005coherent} in the special case $\Gamma_{R_{j}}=\Gamma_{L_{j}}$, corresponding to a symmetric waveguide. 
The net transfer matrix $M$ of the $N$ atom system is given by
\begin{equation}
M=\prod_{j}\mathcal{T}_j :=
\begin{pmatrix}
   M_{11}       & M_{12} \\
    M_{21}      & M_{22}  \\
\end{pmatrix}.
\end{equation}
The net transmission coefficient is given by the formula $T=|t_{N}|^{2}$ and the reflection coefficient is then $R=|r_{1}|^{2}$. Note that $0\leq T\leq 1$ since $0\leq |t_{j}|^{2}\leq 1$ for all $j$. Alternatively, when $\gamma=0$, it can be seen that $T=\vert 1/M_{22}\vert^{2}$, which is a general property of transfer matrices \cite{markos2008wave}.

\section{Band Structure}
In this section we consider the band structure that arises for periodic arrangements of atoms in bidirectional waveguides. We begin by focusing on single photon dispersion properties and then consider the effects of back reflections (deviations from chirality). See~\cite{shen2005coherent, yanik2004stopping,zueco2012microwave} for the case of symmetric waveguides.

\subsection{Dispersion relation}
To study the dispersion characteristics of a single photon, we invoke the periodicity of the infinite lattice and consider solutions of the form
\begin{equation}
\widetilde{t}_{j}=te^{ijKL} \hspace{2mm}\textit{\rm and}\hspace{2mm}\widetilde{r}_{j+1}=re^{ijKL},
\end{equation}
where $K$ is the wavenumber and $L$ is the lattice spacing. By inserting these solutions in (\ref{eq:TransEq}), we find
\begin{equation}
\label{eq:disp2}
\begin{pmatrix}
    t   \\
    r     \\
\end{pmatrix}=e^{-i{K}L}\mathcal{T}
\begin{pmatrix}
    t   \\
    r     \\
\end{pmatrix},
\end{equation}
which means that $e^{i{K}L}$ is an eigenvalue of $\mathcal{T}$. Thus
\begin{equation}
\label{detDR}
\det(\mathcal{T}-e^{i{K}L}\mathcal{I})=0,
\end{equation}
where $\mathcal{I}$ is the $2\times 2$ identity matrix. Eq.(\ref{detDR}) becomes, for $\gamma=0$, the dispersion relation 
\begin{equation}
\cos({K}L)=\frac{\lbrace\Delta^{2}-(\Gamma^{2}_{R}-\Gamma_{L}^{2})\rbrace \cos(q_{R}+q_{L})L/2+2\Delta\Gamma_{R}\sin(q_{R}+q_{L})L/2}{\Delta^{2}+(\Gamma_{R}-\Gamma_{L})^{2}}.
\end{equation}
This result agrees with~\cite{shen2007stopping, zueco2012microwave} for the case of symmetric waveguides with $\Gamma_{R}=\Gamma_{L}$.

\subsection{Small back reflections}
\begin{figure}
\centering
  \begin{tabular}{@{}cccc@{}}
    \includegraphics[width=1.62in, height=1.43in]{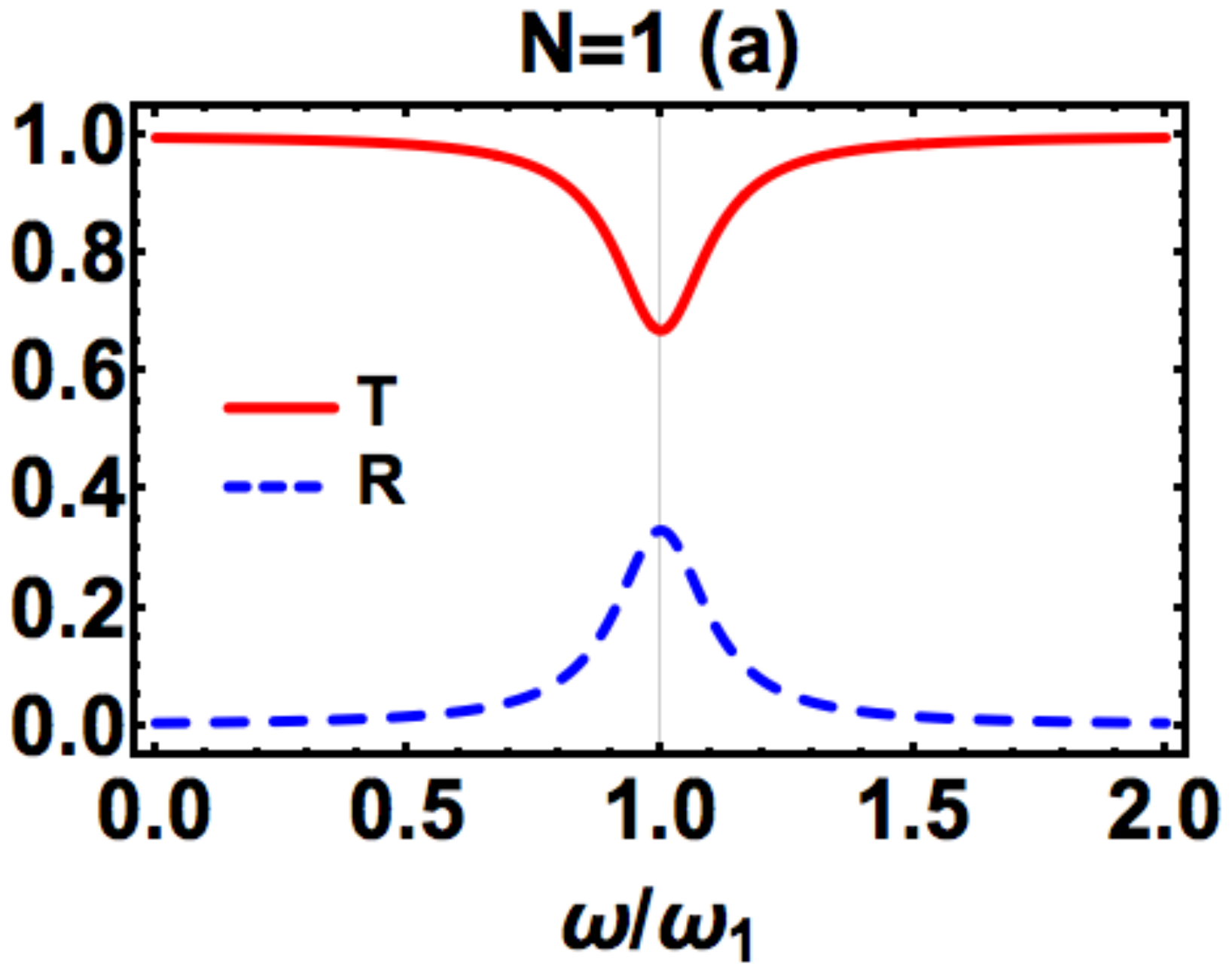} 
  \hspace{-1.5mm}\includegraphics[width=1.68in, height=1.44in]{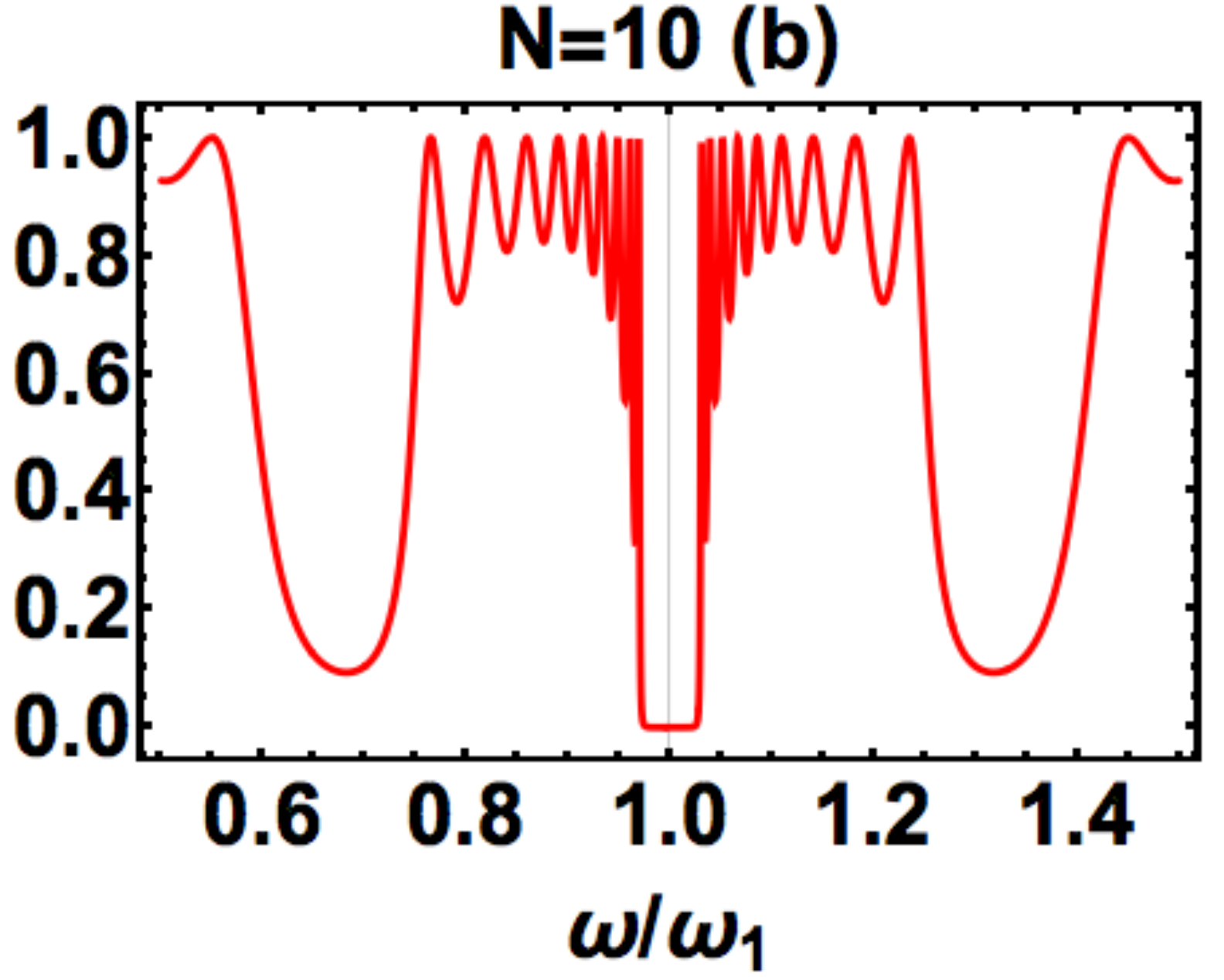}
  \hspace{-2mm} \includegraphics[width=1.68in, height=1.44in]{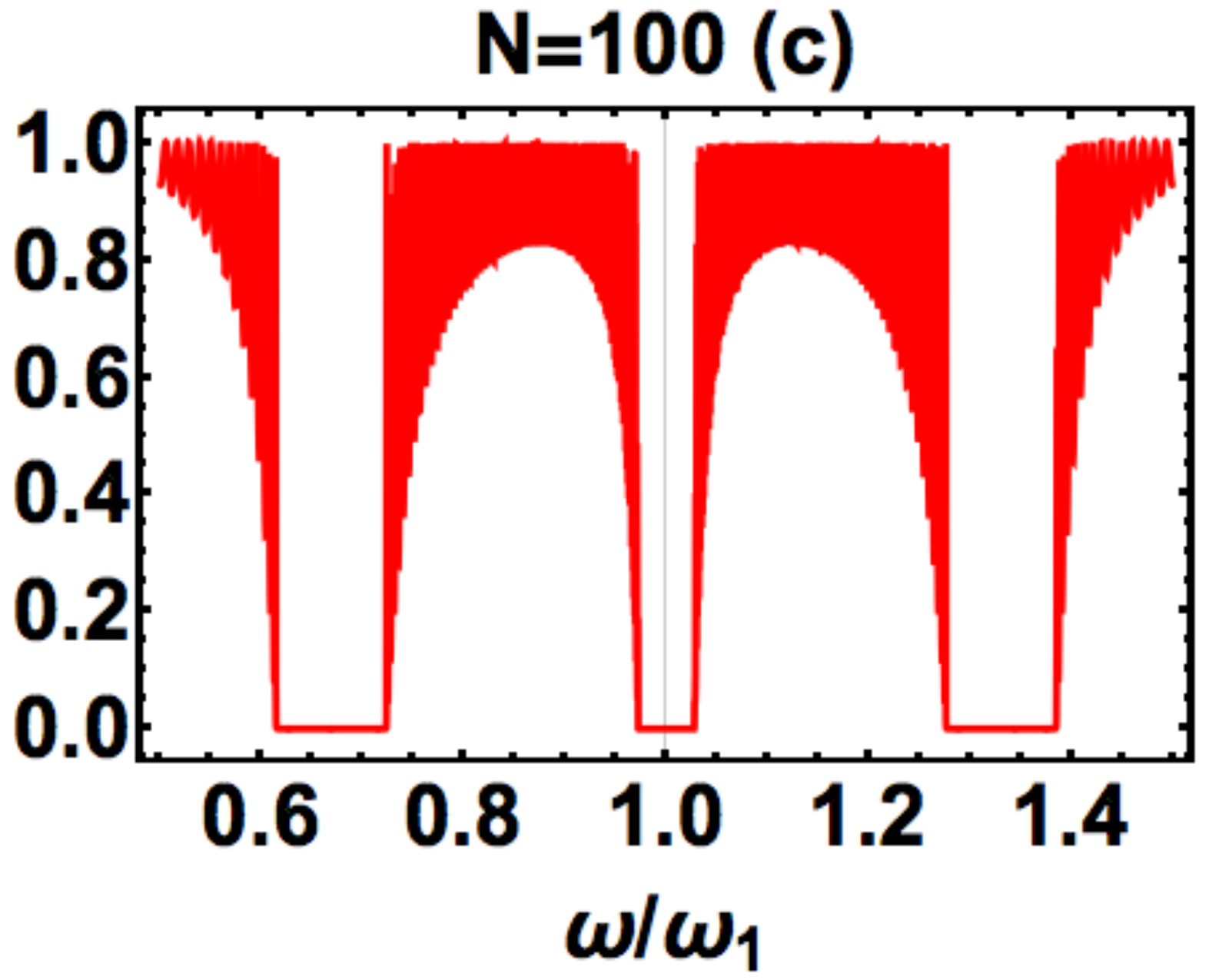}
   \hspace{-1.5mm}\includegraphics[width=1.27in, height=1.7in]{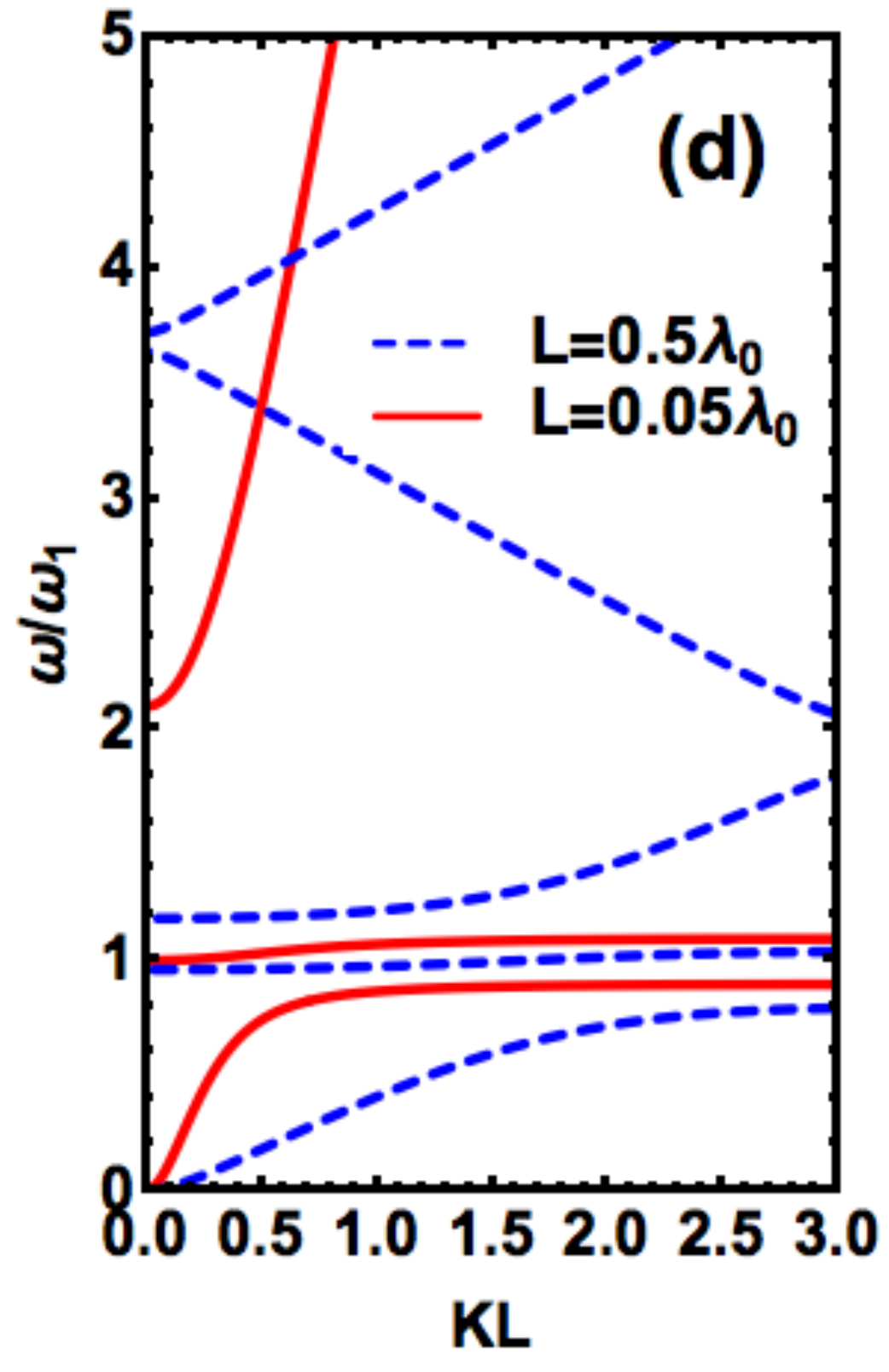}
  \end{tabular}
\captionsetup{
  format=plain,
  margin=1em,
  justification=raggedright,
  singlelinecheck=false
}
 \caption{(Color online) Single-photon transmission and reflection in a waveguide with small back reflections. {(a)} 1, {(b)} 10 and {(c)} 100 periodically arranged identical atoms. The parameters are $\gamma=0,$ $\Gamma_{R}=0.1\omega_{1},$ $v_{L}=10v_{R}$ ($\Gamma_{L}=0.1\Gamma_{R}$) and lattice constant $L=0.5\lambda$ where $\lambda\equiv 2\pi v_{R}/\omega_{1}$. (d) Dispersion curves for two different inter-atomic separations. $\omega_1$ denotes the transition frequency.}\label{Fig5}
\end{figure}

We consider a bidirectional waveguide with small back reflections ($\Gamma_R\gg \Gamma_L$). In Fig.~\ref{Fig5}(a), (b) and (c), we plot the transmission and reflection coefficients $T$ and $R$ as a function of frequency and the number of atoms. For the single-atom case, we find that due to the small atom-waveguide interaction in the backwards (left) direction, $T$ is very large at resonance. However, for the case of multiple atoms a band structure is formed. When $N=100$, the band gaps are clearly visible and the specific range of frequencies on and near resonance where transmission is completely suppressed can be seen.

In Fig.~\ref{Fig5}(d) we plot the dispersion relation for two inter-atomic separations. We observe that larger inter-atomic separations create a higher number of branches. In particular, for larger separations a tiny window of forbidden bands opens up. A similar but wider band gap arises for smaller inter-atomic separations. We see that even a small chiral imbalance can produce sufficient destructive interference to form forbidden bands. Moreover, smaller inter-atomic separations  produce wider gaps.

\subsection{Symmetric waveguides}
\begin{figure}
\centering
  \begin{tabular}{@{}cccc@{}}
  \includegraphics[width=1.62in, height=1.43in]{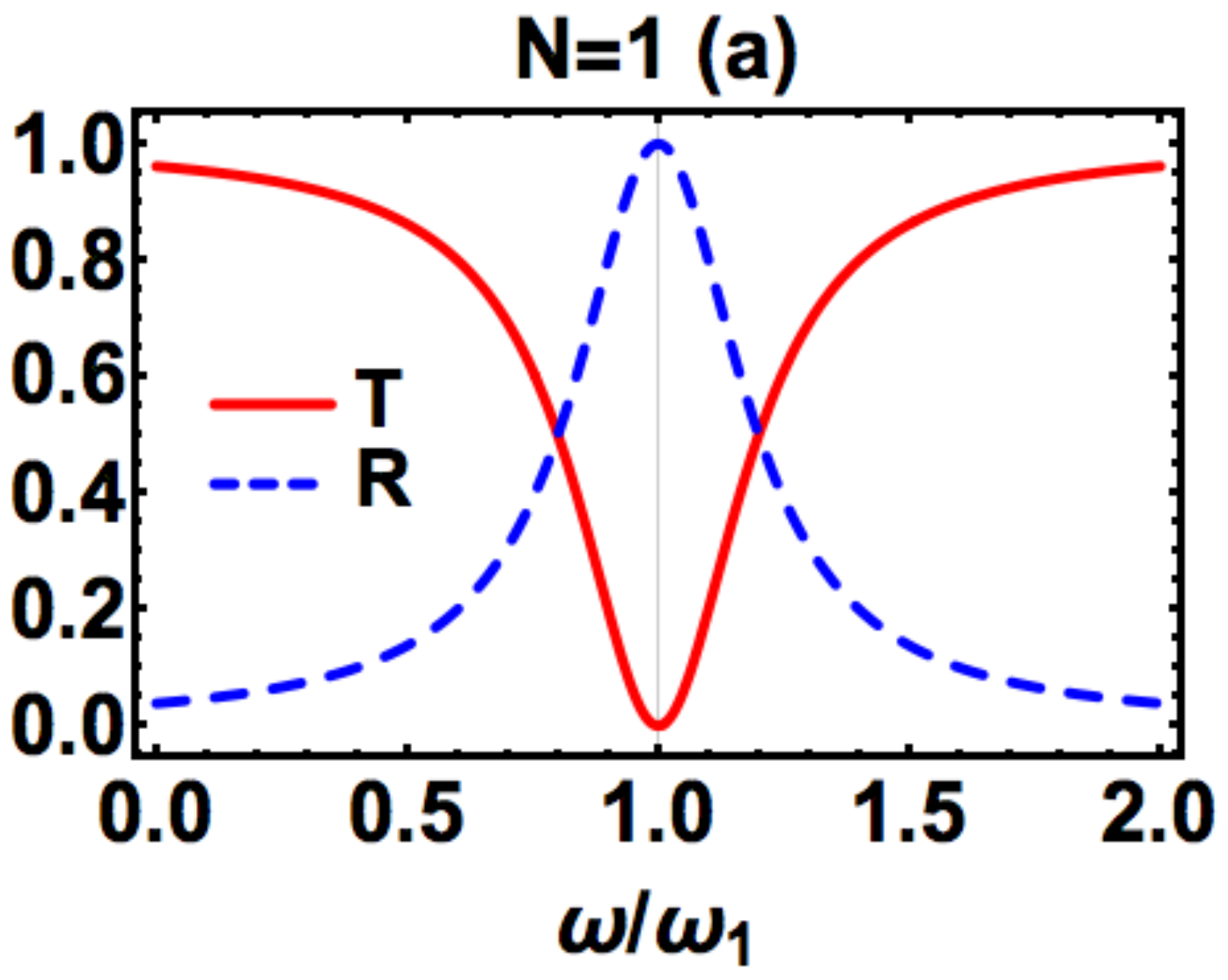} 
  \hspace{-1.5mm}\includegraphics[width=1.68in, height=1.44in]{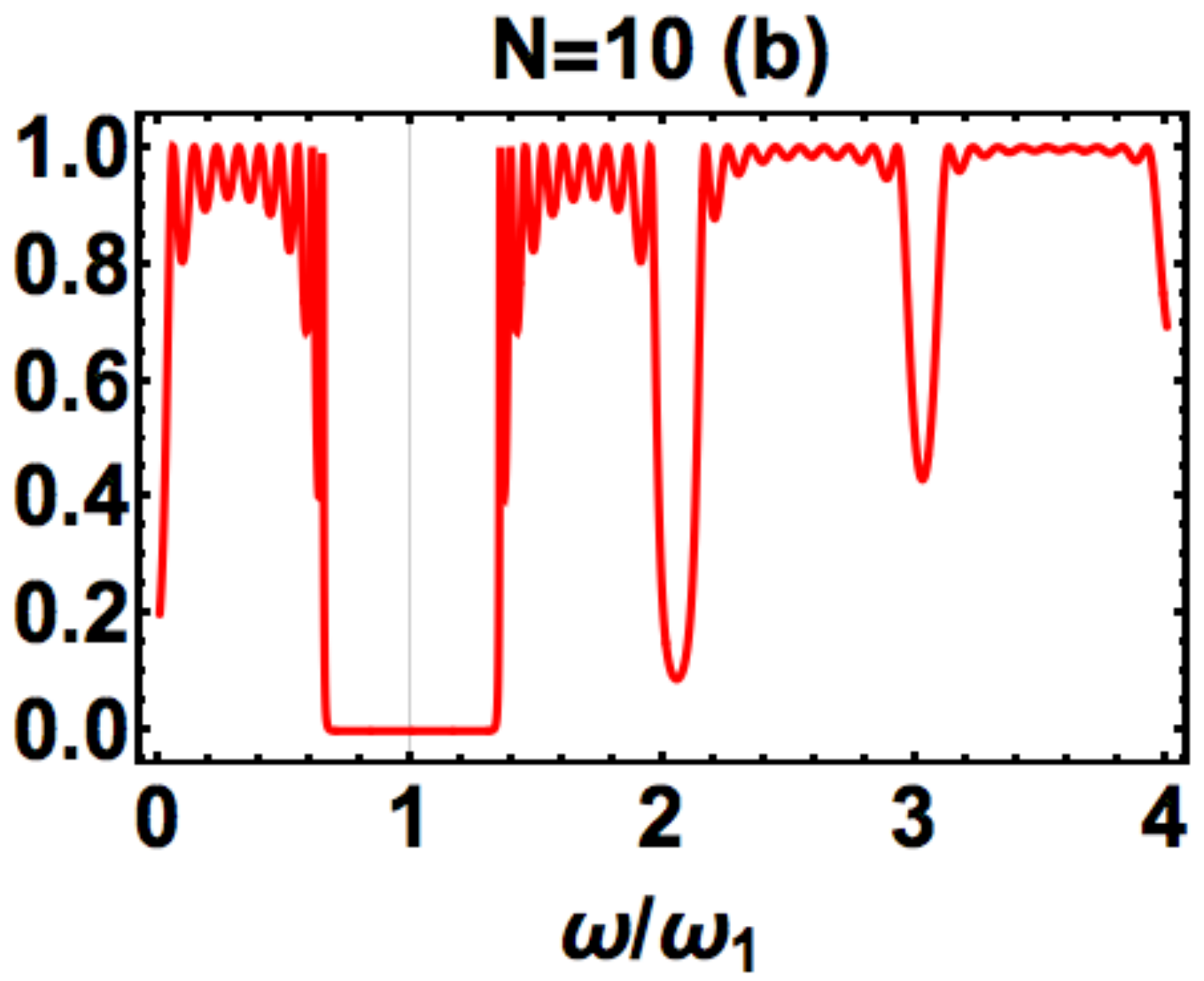}
  \hspace{-2mm} \includegraphics[width=1.68in, height=1.44in]{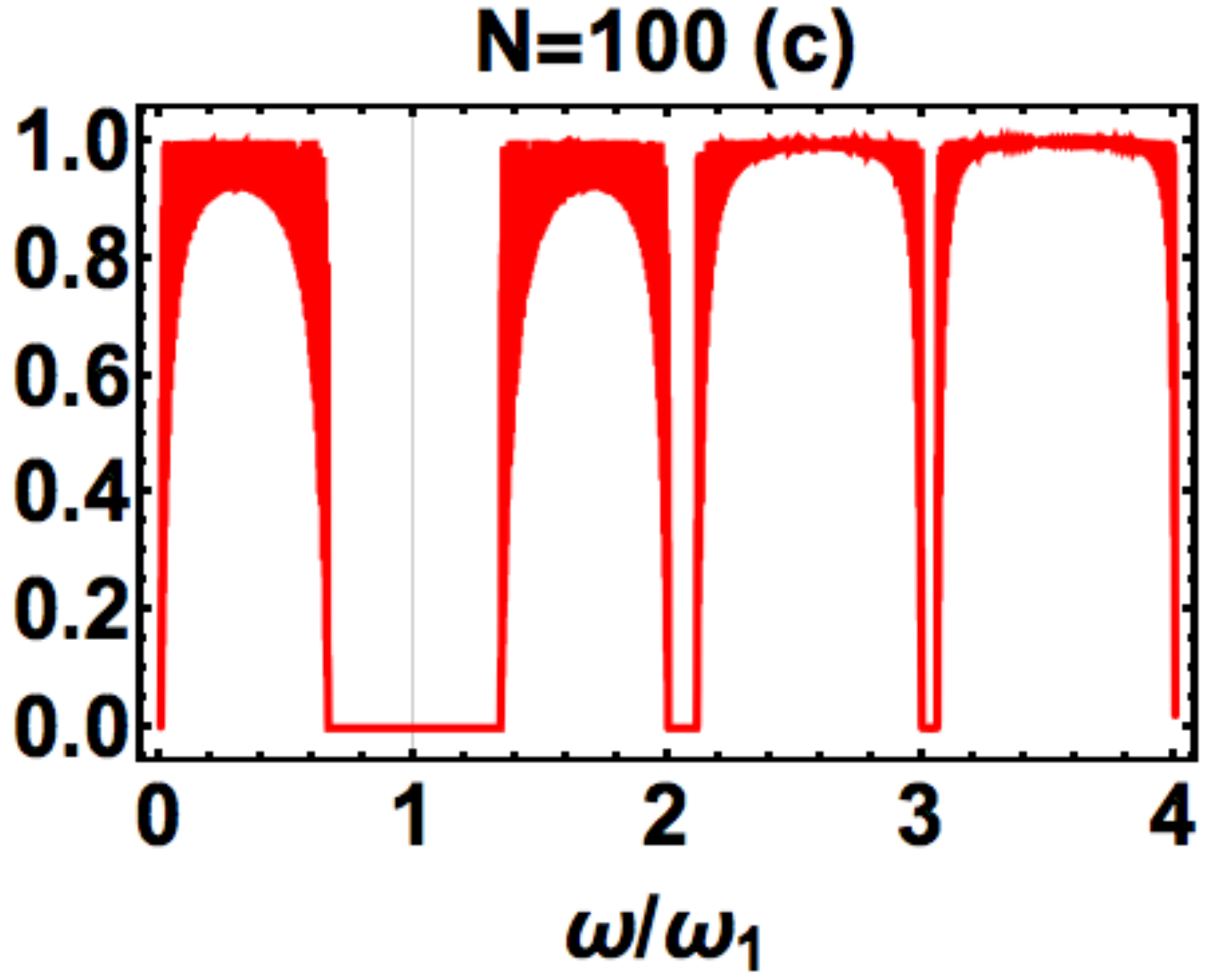}
   \hspace{-1.5mm}\includegraphics[width=1.27in, height=1.7in]{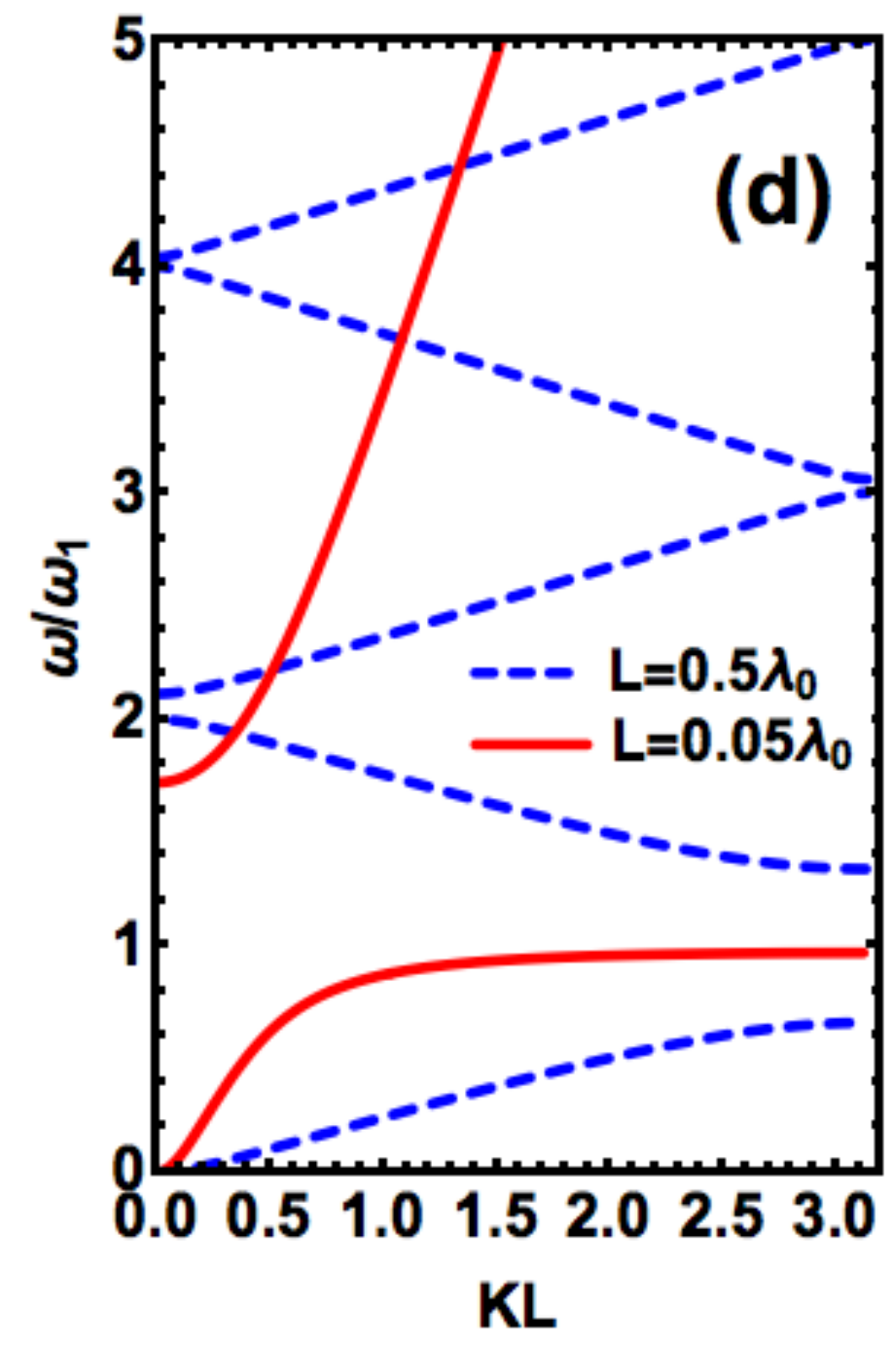} 
  \end{tabular}
\captionsetup{
  format=plain,
  margin=1em,
  justification=raggedright,
  singlelinecheck=false
}
\caption{(Color online) 
Single-photon transmission and reflection in a symmetric waveguide. 
{(a)} 1, {(b)} 10, and {(c)} 100 periodically arranged identical atoms. The parameters are $\gamma=0, \Gamma=0.1\omega_{1}$ and lattice constant $L=0.5\lambda$. (d) Dispersion curves for two inter-atomic separations. $\omega_1$ denotes the transition frequency.}
\label{Fig6}
\end{figure}

We now consider the case of symmetric waveguides with equal group velocities ($\Gamma_{L}=\Gamma_{R}$). In this situation, the atom excitation, transmission and reflection amplitudes for the single atom problem simplify to
\begin{equation}
a=\frac{\sqrt{2v\Gamma}}{{\Delta}+2i\Gamma},\hspace{2mm} t=\frac{{\Delta}}{{\Delta}+2i\Gamma}, \hspace{2mm} r=\frac{-2i\Gamma}{{\Delta}+2i\Gamma}.
\end{equation} 
In Fig.~\ref{Fig6} we plot the transport properties of the system. In the single atom case, the net reflection coefficient manifests a Lorentzian profile with unit value of $R$ at resonance. In the multiple atom scenarios, a full photonic band gap emerges, which allows for the possibility of generating frequency comb patterns \cite{liao2016single}. However, as compared to the case of small back reflections, the width of the gap is larger on resonance. For $N=10$, the off-resonance bands appear as thin peaks with decreasing heights as we move away from resonance.
In Fig.~\ref{Fig6}(d) we show the dispersion curves. The dispersion relation becomes
\begin{equation}
\cos({K}L)=\cos(qL)+\frac{2\Gamma}{\Delta}\sin(qL) ,
\end{equation}
where $q_R=q_L=q$. Similar to the small back reflection case, we note that the band gap structure can be engineered by altering the separation between the atoms. Beyond this general feature, in the symmetric case, the width and the locations of the band gaps are considerably changed.  For instance, the tiny forbidden gap appearing at $\sim 1.2\omega_{1}$ for small back reflections  has been shifted to $\sim 2\omega_{1}$ for the symmetric problem, where $\omega_1$ is the transition frequency. Moreover, when we compare Fig.~\ref{Fig5}(d) and Fig.~\ref{Fig6}(d), we see that the dispersion curve appearing at $\omega=\omega_{1}$ with $KL\lesssim 0.5$ for the small back reflection problem does not appear. 

\section{Effects of Disorder}
\subsection{Evidence for localization}

For chiral waveguides, we were able to establish the existence of localization and calculate the localization length analytically. However, for bidirectional waveguides an analysis along the same lines is not straightforward. Instead, we seek numerical evidence for localization by following (\ref{llc}) and plotting $\langle \ln T\rangle$ as a function of the number atoms $N$. In Fig.~\ref{Fig7} we consider four cases that we will study in detail in later sections: position disorder in symmetric waveguides or with small back reflections, and frequency disorder in symmetric waveguides or with small back reflections. We see that $\langle \ln T\rangle$ decreases linearly with $N$, consistent with (\ref{llc}). Based on this result, we compute the localization length according to (\ref{llc}).

\begin{figure}[t]
\centering
  \begin{tabular}{@{}cccc@{}}
\includegraphics[width=2.25in, height=1.5in]{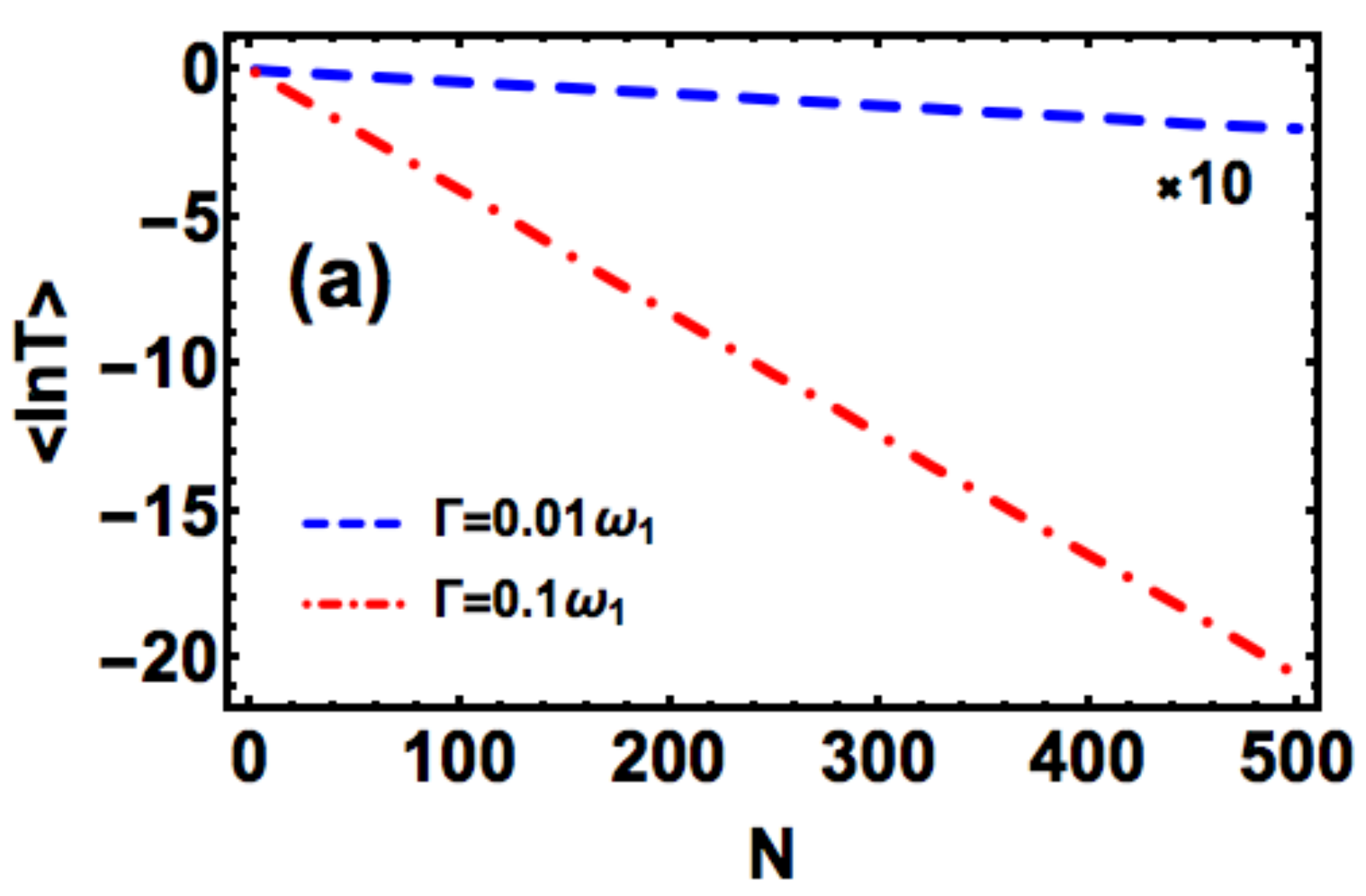} 
  \hspace{-1mm}\includegraphics[width=2.17in, height=1.48in]{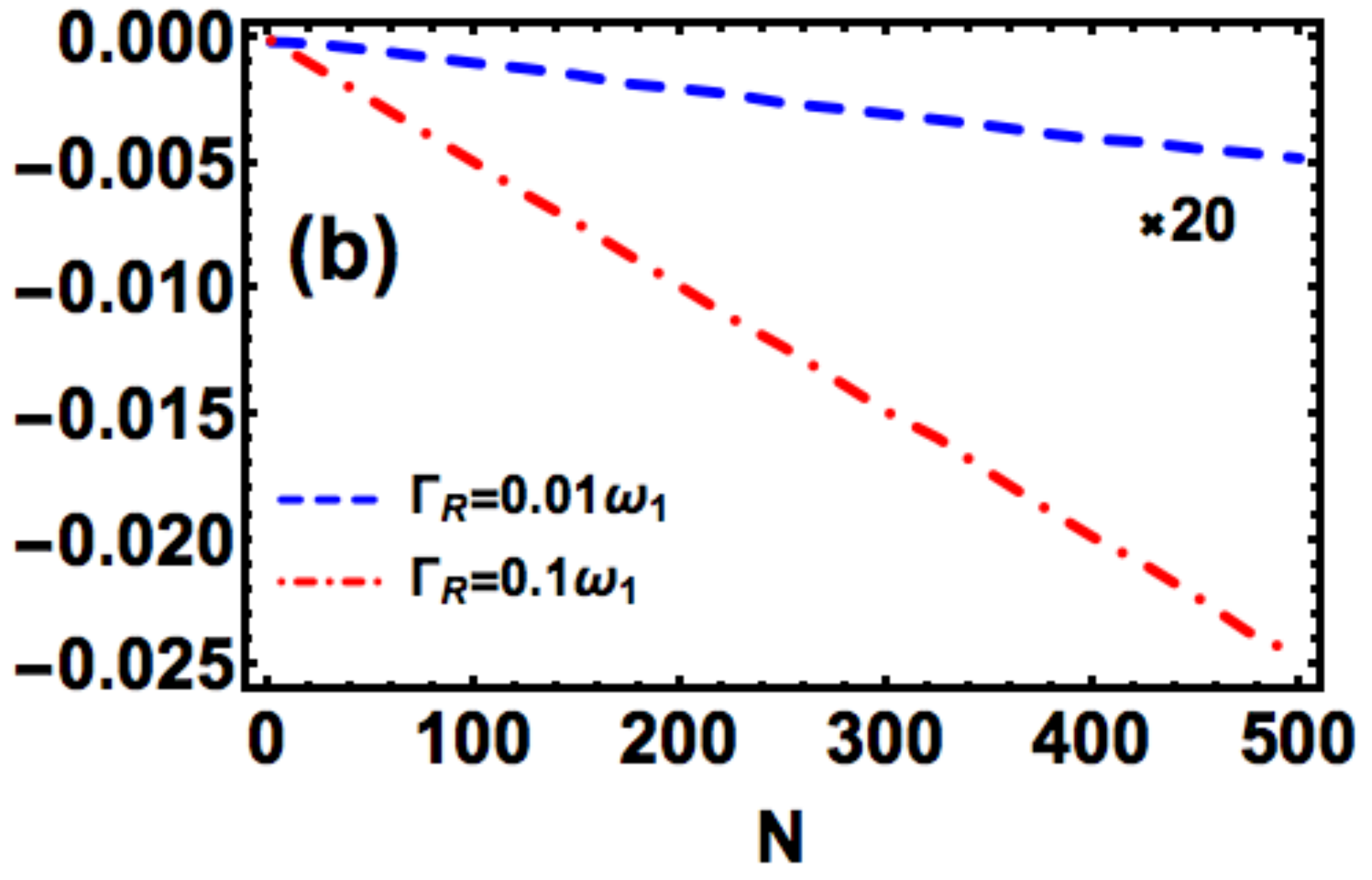}
   \hspace{-1mm}\includegraphics[width=2.15in, height=1.5in]{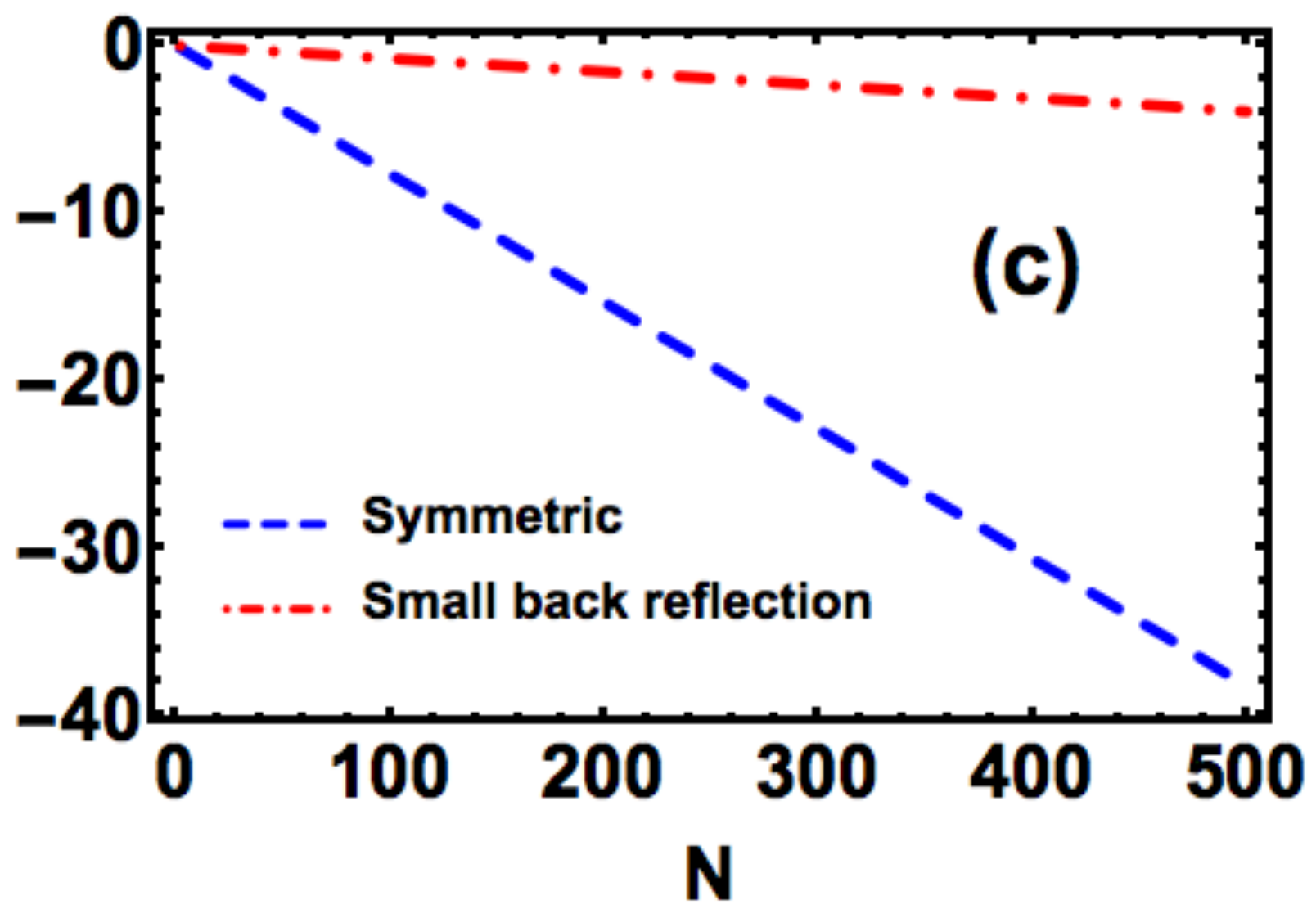} 
  \end{tabular}
\captionsetup{
  format=plain,
  margin=1em,
  justification=raggedright,
  singlelinecheck=false
}
\caption{(Color online) 
Dependence of $\langle \ln T \rangle$ on the number of atoms $N$ for position disorder. Here $\omega=2\omega_{1}$ for (a) symmetric waveguide and (b) small back reflections with $\Gamma_{L}=0.1\Gamma_{R}$ is considered. The mean interatomic separation is $\lambda/2$ and the strength of the disorder $\sigma=\lambda$.
(c) Frequency disorder. A periodic chain of atoms is considered with a lattice constant $L=\lambda/2$. The strength of the disorder for small back reflections (symmetric waveguides) is $\sigma=\Gamma_{R}$ ($\Gamma$) and the mean is $2\Gamma_{R}$ ($2\Gamma$). 
We have set $\gamma=0$ (no spontaneous decay)  and performed the average over $10^{4}$ realizations of the disorder. The error bars are too small to be displayed.}
\label{Fig7}
\end{figure}

\subsection{Small back reflections}
\subsubsection{Position disorder}

We begin by considering a position-disordered 10 atom chain. In Fig.~\ref{Fig8}(a), we see that the band structure observed in the corresponding periodic setting has disappeared (for comparison see Fig.~\ref{Fig5}(b)). In addition, on resonance a small region of minimal transmission forms. In Fig.~\ref{Fig8}(b) we plot the localization length $\xi$ as a function of frequency $\omega$. We find that $\xi$ reaches its minimum value at resonance, where the system is almost completely reflecting. Far from resonance, $\xi$ is considerably enhanced due to the increased transmission. These trends suggest the possibility of forming frequency-dependent localized states due to small back reflections. 

\begin{figure}
\centering
   \begin{tabular}{@{}cccc@{}}
   \includegraphics[width=2.5in, height=1.75in]{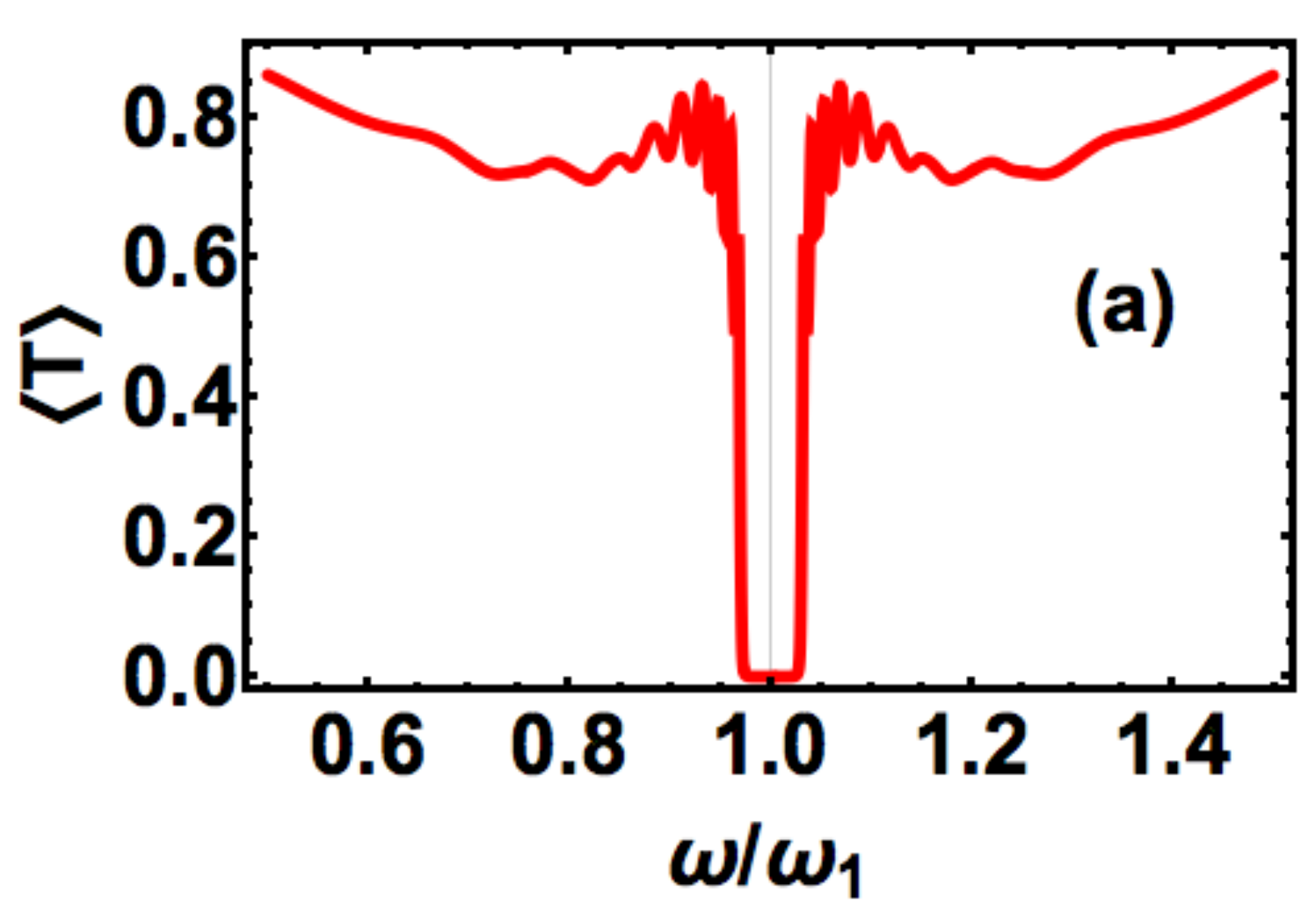} 
  \hspace{-1mm}\includegraphics[width=2.45in, height=1.75in]{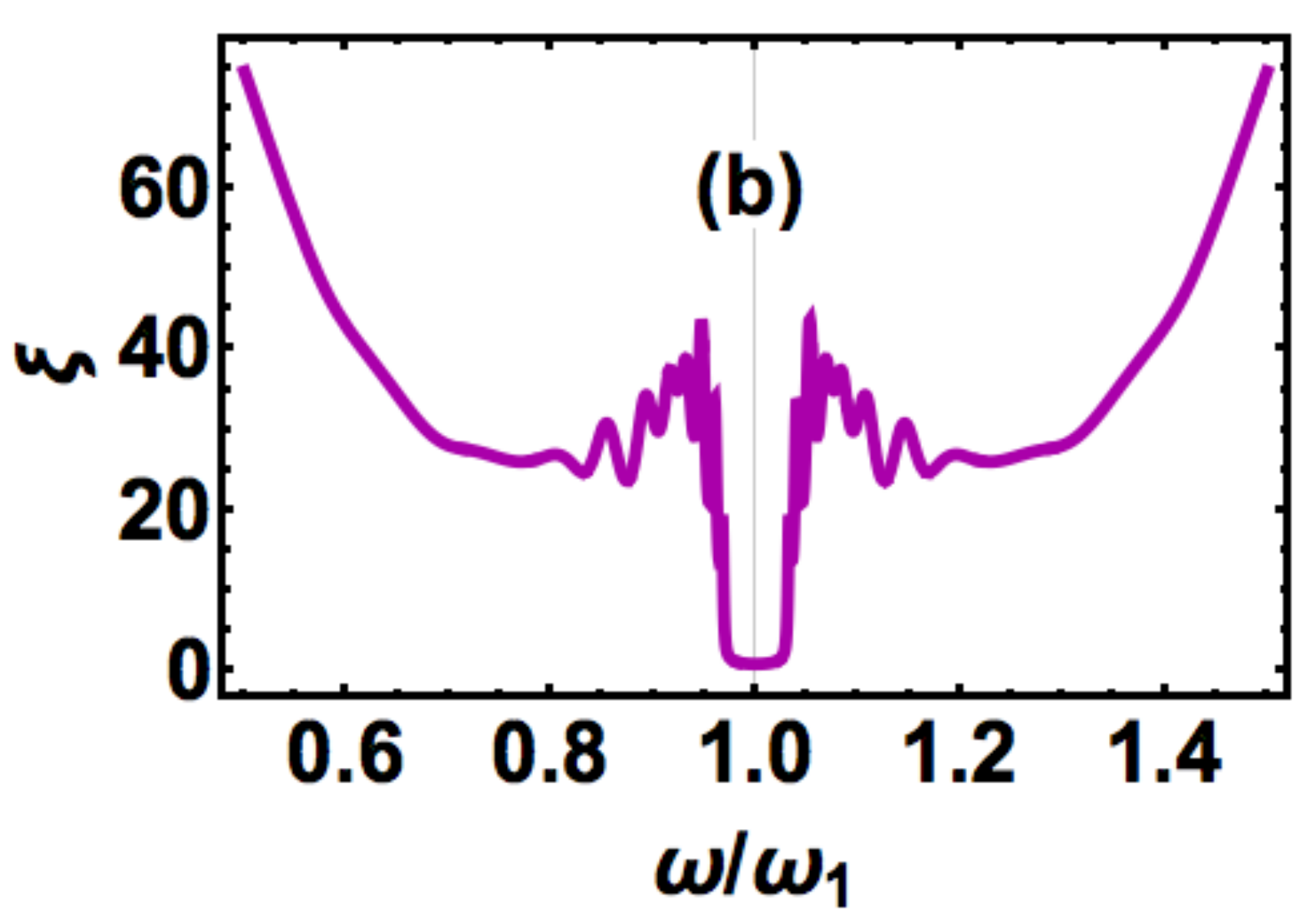}\\
   \hspace{-1mm} \includegraphics[width=2.48in, height=1.8in]{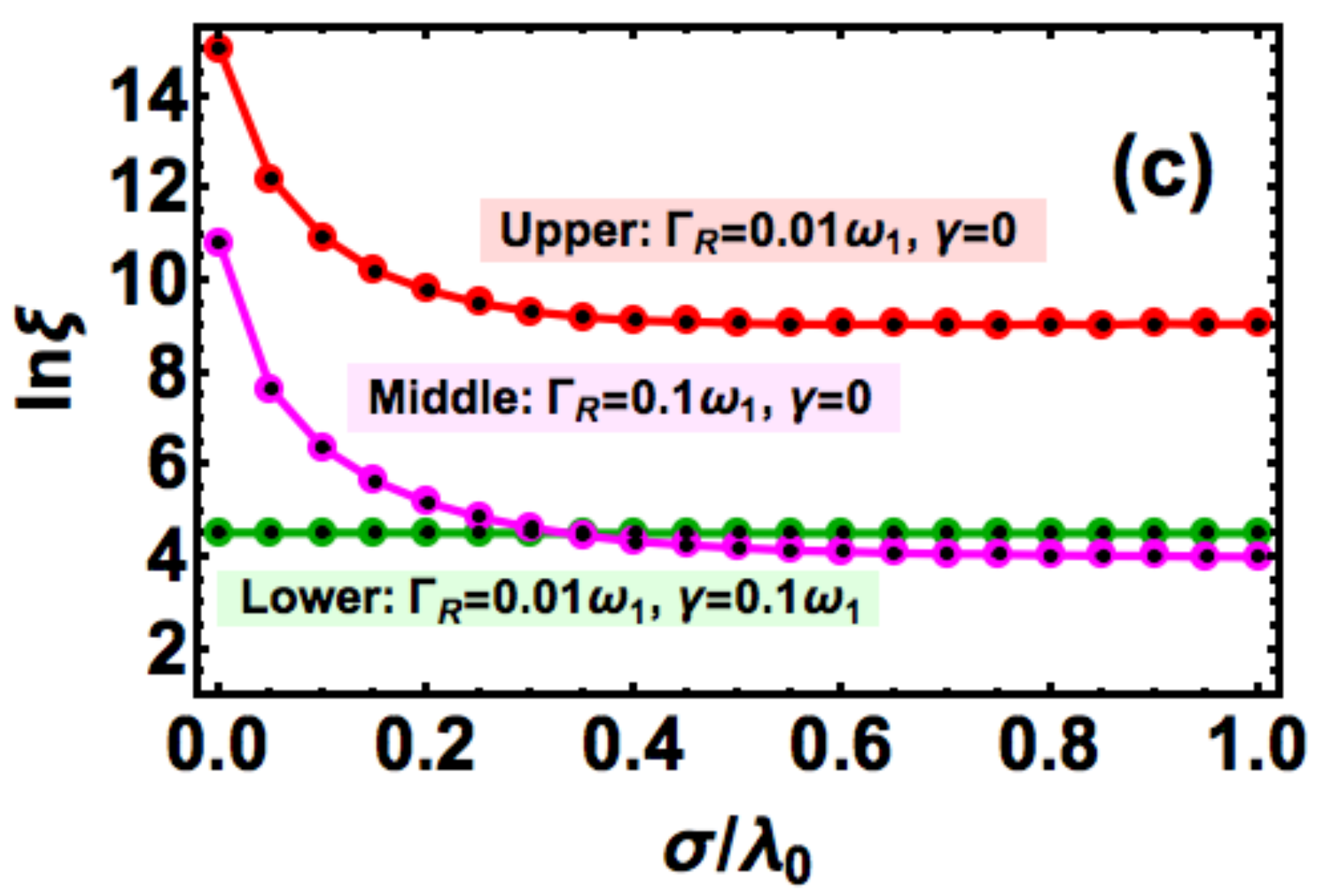}
     \hspace{-3mm} \includegraphics[width=2.48in, height=1.8in]{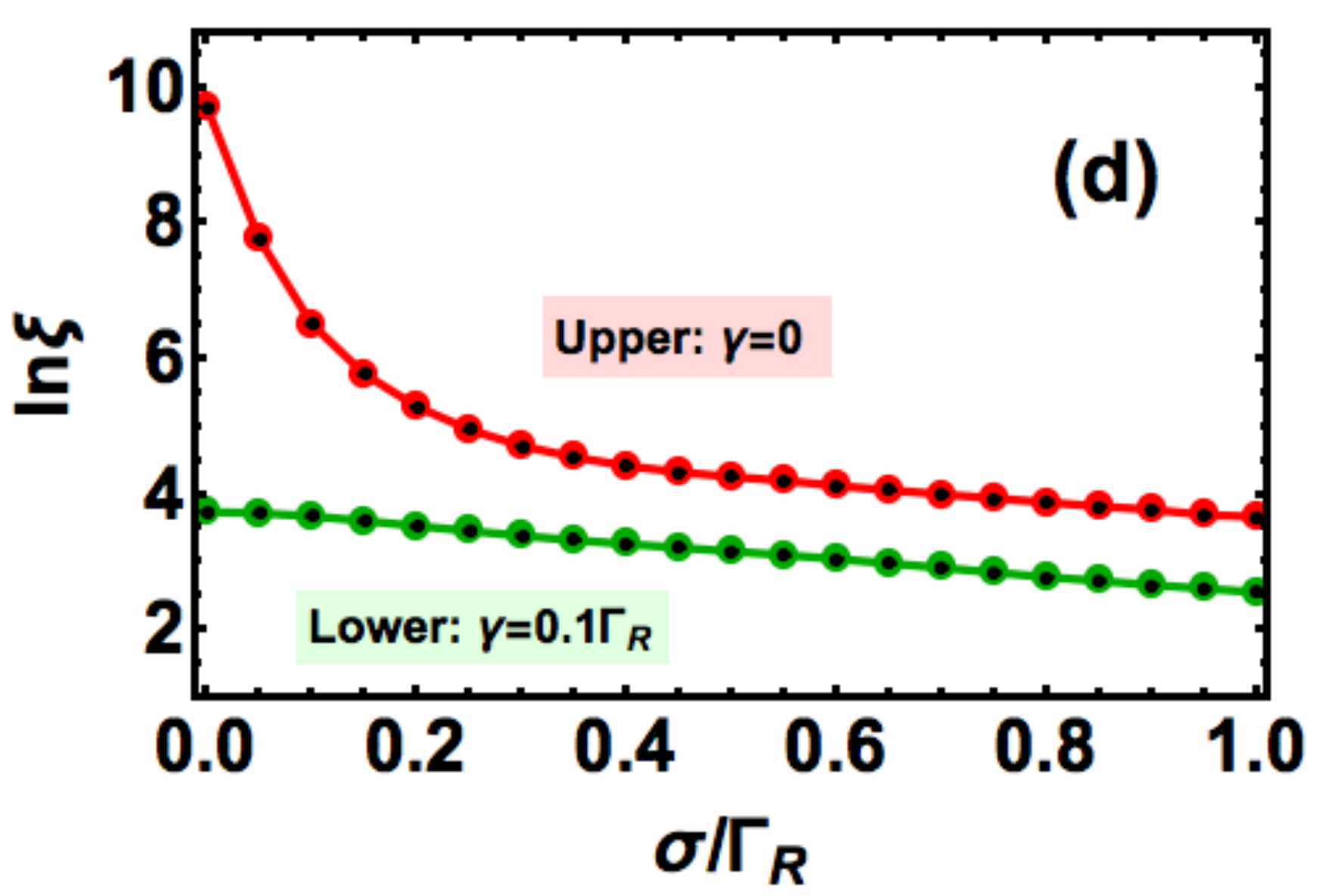}
  \end{tabular} 
\captionsetup{
  format=plain,
  margin=1em,
  justification=raggedright,
  singlelinecheck=false
}
\caption{(Color online) Localization with small back reflections. In all plots $v_{L}=10v_{R}$ (equivalently $\Gamma_{L}=0.1\Gamma_{R}$). In (a) and (b) $\gamma=0$, $N=10$, the mean spacing is $\lambda/2$ and the disorder strength $\sigma=2\lambda $.  The average is performed over 500 realizations of the disorder. 
(a) Average transmission $\langle T \rangle$ as a function of frequency $\omega$ for position disorder. (b) Localization length $\xi$ versus $\omega$ for position disorder. (c) Localization length $\xi$ versus $\sigma$ for position disorder with mean spacing $\lambda/2$, $\omega=1.6\omega_{1}$, $N=10^3$ and $10^4$ realizations. (d) $\xi$ versus $\sigma$ for frequency disorder with atomic spacing $\lambda/2$, mean frequency of $3\Gamma_R$, $N=10^3$ and $10^4$ realizations.
}
\label{Fig8}
\end{figure}

In Fig.~\ref{Fig8}(c) we consider a position-disordered chain with $N=10^{3}$ atoms. We plot the dependence of the localization length $\xi$ on the strength of disorder $\sigma$. We assume that the system is tuned away from resonance (to allow transmission) and consider the cases of weak and strong coupling of the atoms to the waveguide. We find that $\xi$ is a decreasing function of $\sigma$ and that $\xi$ is smaller for strong coupling. We also explore the effect of spontaneous emission on $\xi$. As expected, we find that spontaneous emission is the dominant mechanism to destroy photon transport and the dependence of $\xi$ on $\sigma$ is very weak.

\subsubsection{Frequency disorder}
In Fig.~\ref{Fig8}(d) we plot $\xi$ versus $\sigma$ for frequency disorder. The system is taken far from resonance and shows large transmission for small disorder with $\gamma=0$. For $\gamma\neq 0$ the transmission remains very small for all values of $\sigma$. The overall behavior is similar to that of position disorder.

\subsection{Symmetric waveguides}
%\subsubsection{Position disorder}

We now consider the case of symmetric waveguides, following along the same lines as the discussion of waveguides with small back reflections. The results are presented in Fig.~\ref{Fig9}. We see that the behavior of the transmission and localization length mirrors that in Fig.~\ref{Fig8}. However, it is important to note that for
symmetric waveguides, the scale of $\xi$ is decreased by an order of magnitude compared to waveguides with small back reflections.
We also note that if the frequency of the photon $\omega$ lies in a bandgap of the corresponding periodic system, then the dependence of the localization length on the strength of the disorder is generally not decreasing (data not shown).

\begin{figure}
\centering
 \begin{tabular}{@{}cccc@{}}
   \includegraphics[width=2.5in, height=1.75in]{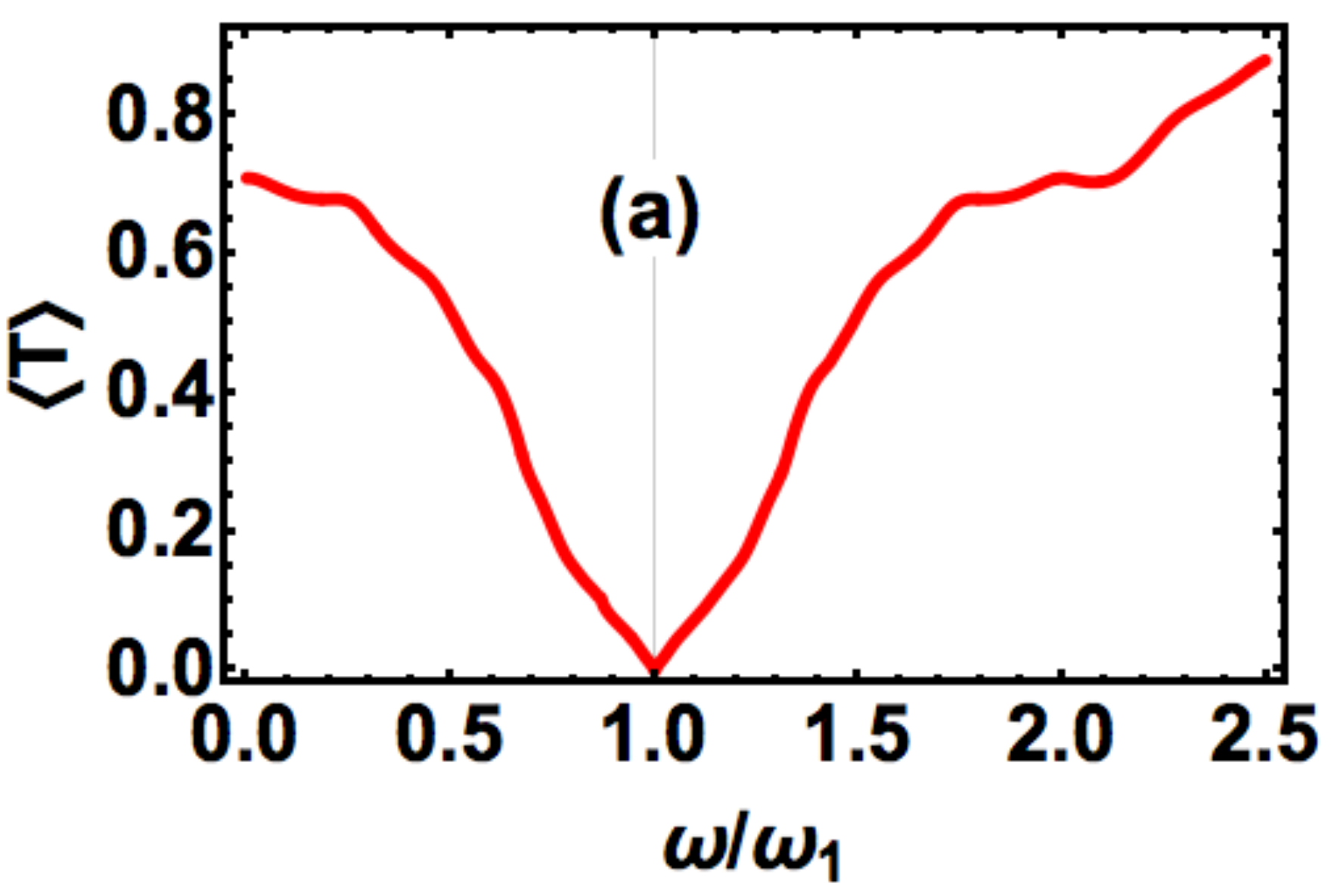} 
  \hspace{-1mm}\includegraphics[width=2.45in, height=1.75in]{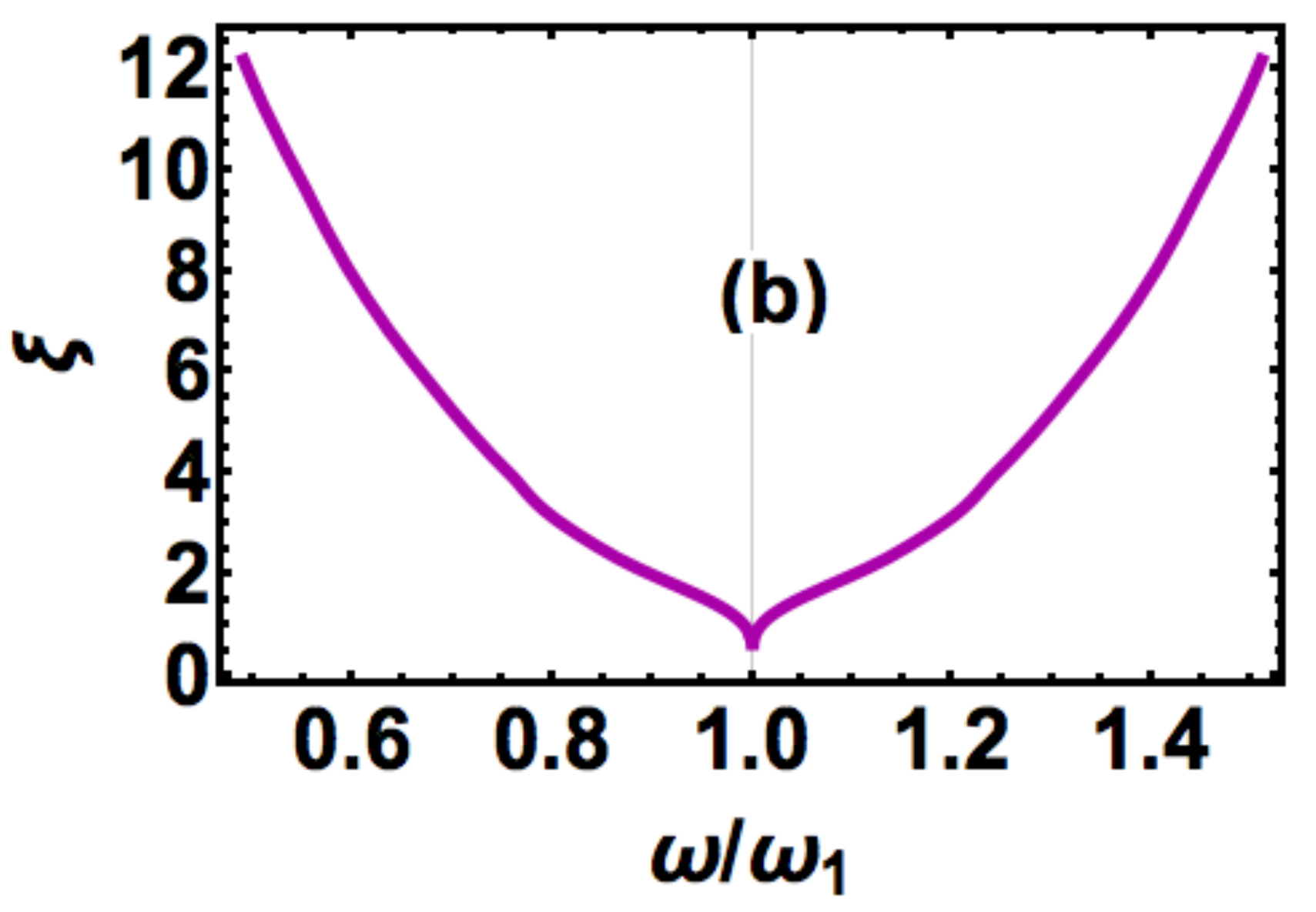}\\
   \hspace{-1mm} \includegraphics[width=2.48in, height=1.8in]{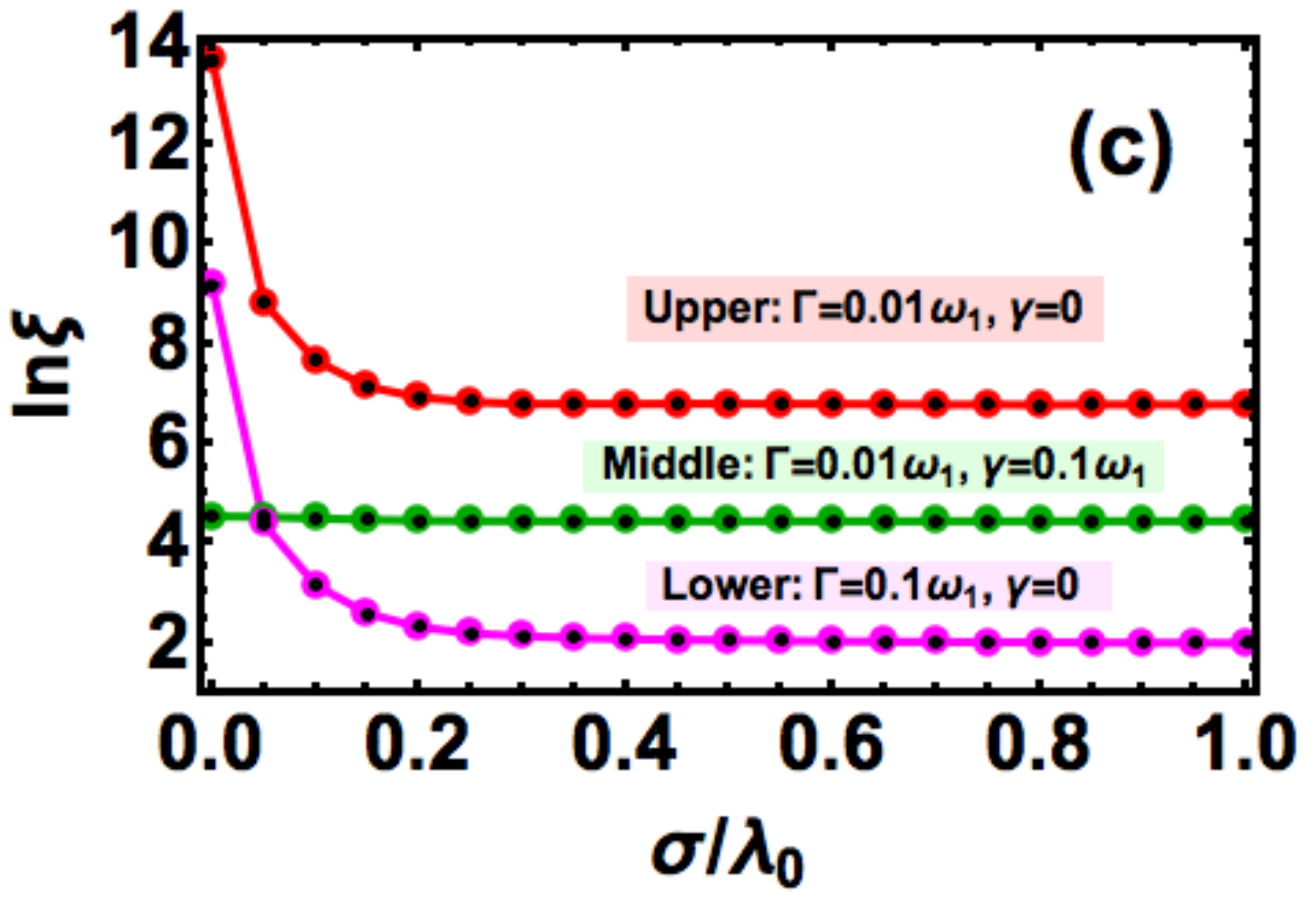}
     \hspace{-3mm} \includegraphics[width=2.48in, height=1.8in]{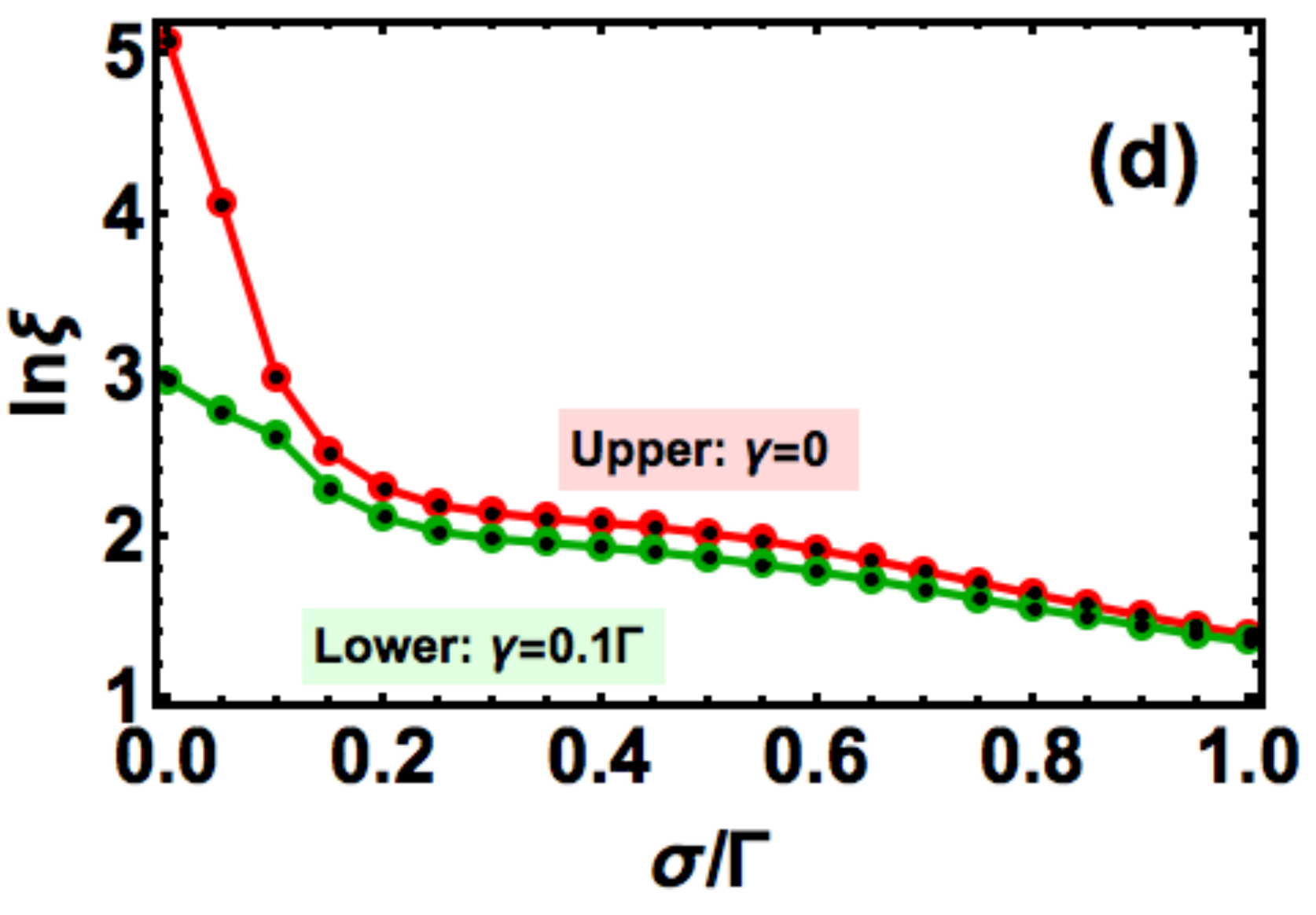}
  \end{tabular}
\captionsetup{
  format=plain,
  margin=1em,
  justification=raggedright,
  singlelinecheck=false
}
\caption{(Color online) Localization in a symmetric waveguide. The parameters are the same as in Fig.~\ref{Fig8}. }
\label{Fig9}
\end{figure}

\section{Discussion}
We have investigated the problem of single-photon transport in chiral and non-chiral waveguide QED. We have considered the band structure that arises from periodically arranged atoms and have studied the effects of disorder in atomic positions and transition frequencies. Our conclusions may be summarized as follows.  

The absence of backscattering in chiral waveguides precludes the existence of band structure in periodic systems. In addition, chiral systems are immune to position disorder and do not exhibit localization. However, localization does arise in chiral waveguides with frequency disorder, a setting in which it is possible to calculate the average transmission and localization length analytically.

Bidirectional waveguides generally exhibit a band structure for periodic systems. The width and location of the bands is controlled by the symmetry of the waveguide. We have found that both positional and frequency disorder lead to localization and that the localization length takes the smallest value at resonance for both types of bidirectional waveguides. For position disorder, strong atom-waveguide coupling generally leads to smaller localization lengths compared to systems with weak coupling.

%The results presented in this study have applications in quantum information processing, where novel types of photon storage devices based on disordered nano-photonic systems are considered. In addition, applications to photonic analogs of many-body condensed matter systems  can be envisioned.
 
\acknowledgments

This work was supported in part by the NSF grants DMR-1120923 and DMS-1619907.

\bibliographystyle{ieeetr}
\bibliography{Paper}
\end{document}